\documentclass{elsart1p}
\usepackage{graphicx}
\usepackage{amssymb}
\usepackage{rotating}
\usepackage{subfigure}

\newcommand{\gvec}[1]{\mbox{$\bf #1$}}

\newcommand{\beq}{\begin{equation}}
\newcommand{\eeq}{\end{equation}}
\newcommand{\beqa}{\begin{eqnarray}}
\newcommand{\eeqa}{\end{eqnarray}}

\newcommand{\lp}{\left(}
\newcommand{\rp}{\right)}
\newcommand{\Wi}{\mbox{$W_i$}}

\newcommand{\Wij}{\mbox{W$_{ij}$}}

\newcommand{\url}[1]{{\tt http://#1}}
\newcommand{\DE}{\mbox{$\Delta E^{\mbox{\tiny thermal}}$}}
\newcommand{\DEij}{\mbox{$\Delta E_{ij}^{\mbox{\tiny thermal}}$}}
\newcommand{\sgn}[1]{\mbox{sgn$( #1 )$}}
\newcommand{\fij}{\mbox{$f_{ij}$}}
\newcommand{\fijsm}{\mbox{$f_{ij}^{\mbox{\tiny sm}}$}}
\newcommand{\fijmono}{\mbox{$f_{ij}^{\mbox{\tiny mono}}$}}

\begin{document}
\begin{frontmatter}

\title{A compatibly differenced total energy conserving form of SPH}
\author{J. Michael Owen}
\address{Lawrence Livermore National Laboratory, M/S L-38, P.O. Box 808, Livermore, CA, 94550}
\ead{mikeowen@llnl.gov}

\begin{abstract}
We describe a modified form of Smoothed Particle Hydrodynamics (SPH) in which
the specific thermal energy equation is based on a compatibly differenced
formalism, guaranteeing exact conservation of the total energy.  We compare the
errors and convergence rates of the standard and compatible SPH formalisms on
analytic test problems involving shocks.  We find that the new compatible
formalism reliably achieves the expected first-order convergence in such tests,
and in all cases improves the accuracy of the numerical solution over the
standard formalism.
\end{abstract}

\end{frontmatter}

\section{Introduction}
Smoothed Particle Hydrodynamics (SPH) is a meshless or particle based Lagrangian
approach to modeling hydrodynamics.  SPH was originally developed as a method
for studying astrophysical systems (\cite{Lucy77}, \cite{GM77}), for which
purpose it has a number of advantages: it's particle-like nature is well suited
for combination with existing N-body gravitational techniques; the Lagrangian
frame of SPH naturally follows the large dynamic range of length and mass scales
that gravitationally unstable systems undergo; the lack of an imposed geometry
from an underlying mesh suits the shifting, complex, three-dimensional nature of
astrophysical objects.  These properties have made SPH modeling a successful
tool for studying galaxy formation and evolution, star formation, the formation
and evolution of planetary bodies, etc.  SPH has also found many applications
outside of astrophysics as well, such as modeling of metal casting, material
fracture, impact of bodies into water, and multi-phase flows, among others.


The standard SPH formalism (outlined in \cite{Mreview92} \& \cite{Breview90})
time evolves the specific thermal energy in a manner that does not conserve
total energy, and in recent years it has been found that the energy conservation
with this approach can become quite poor (\cite{Hernquist93},
\cite{NelsonPap94}, \cite{EntropySPH02}).  These authors suggest a few
modifications on the standard SPH formalism to improve the energy conservation:
\cite{NelsonPap94} and \cite{PriceMon06} find that accounting for the
variability of the SPH smoothing scale in the SPH dynamical equations improves
the consistency and thereby the energy conservation of the scheme; in
\cite{EntropySPH02} the authors choose to evolve the specific entropy equation
in place of the energy equation (including a term which also accounts for the
variable smoothing scale) and likewise find that overall energy conservation is
improved.  None of these approaches guarantees energy conservation, however,
typically improving the total conservation to $\sim$1\% on tests cited.

In this paper we adopt a different approach to this problem, and formulate an
exactly energy preserving form of SPH based on the concept of compatible
differencing outlined in \cite{CompatibleHydro98}.  The resulting form of SPH is
manifestly energy preserving (to machine roundoff) by construction, irrespective
of details such as the presence of a variable smoothing scale.  This paper is
layed out as follows: \S \ref{standard.sec} describes the standard SPH formalism
we will be comparing with; \S \ref{compatible.sec} derives the new compatible
method for evolving the SPH specific thermal energy; \S \ref{tests.sec}
demonstrates the performance of the new scheme on several test problems of
interest; finally \S \ref{conclusions.sec} discusses our conclusions.

\section{The standard SPH equations}
\label{standard.sec}
Before we begin, a word about notation.  In this paper we use the convention
that Latin subscripts denote node indices ($m_i$ is the mass of node $i$),
while Greek superscripts denote dimensional indices ($x_i^\alpha$ is the
$\alpha$th component of the position for node $i$).  We use the summation
convention for repeated Greek indices: $x_i^\alpha v_i^\alpha = \gvec{x}_i \cdot
\gvec{v}_i$.  For conciseness we represent the spatial gradient as
$\partial_\alpha F \equiv \partial F/\partial x^\alpha$.

The SPH dynamical equations describing the evolution of the mass, momentum, and
energy have been derived elsewhere in detail (see for instance \cite{Mreview92}
and \cite{Breview90}), so we will simply summarize the forms we are using:
\beq
  \label{sphmass.eq}
  \rho_i = \sum_j m_j \Wi,
\eeq
\beq
  \label{tsphmom.eq}
  \frac{D v_i^\alpha}{Dt} =
	 -\sum_j m_j \left[ \lp \frac{P_i}{\rho_i^2} + \frac{P_j}{\rho_j^2} \rp
                            \partial_\alpha \Wij +
                            \Pi^{\beta \alpha}_{ij}
                            \partial_\beta \Wij
	             \right],
\eeq
\beq
  \label{tsphenergy.eq}
  \frac{Du_i}{Dt} = \sum_j m_j \lp \frac{P_i}{\rho_i^2}
                    v_{ij}^\alpha \partial_\alpha \Wij +
	            \frac{1}{2} \Pi^{\alpha \beta}_{ij}
                    v_{ij}^\beta \partial_\alpha \Wij
                    \rp,
\eeq
where for a given node $i$ $\rho_i$ is the mass density, $m_i$ the mass,
$v^\alpha_i$ the velocity, $P_i$ the pressure, and $u_i$ the specific thermal
energy.  $W$ and $\partial_\alpha W$ represent the interpolation kernel and it's
gradient -- the form of $W$ we use is the cubic B-spline, given in
\cite{Mreview92}.  Because $W(r, h)$ is a function of distance and the smoothing
scale, the subscripts on $W_i = W(r, h_i)$ and $W_j = W(r, h_j)$ denote which
nodes definition of the smoothing scale has been used.  The terms with both $i$
and $j$ indices indicate either differences ($v_{ij}^\alpha = v_i^\alpha -
v_j^\alpha$) or explicitly symmetrized quantities: $W_{ij} = (W_i + W_j)/2$;
$\partial_\alpha W_{ij} = (\partial_\alpha W_i + \partial_\alpha W_j)/2$.

The term $\Pi^{\alpha \beta}_{ij}$ is the artificial viscosity, expressed here
appropriately for a tensor viscosity as described in \cite{TensorQ}.  The
majority of tests presented in this paper are one-dimensional, in which case
$\Pi^{\alpha \beta}_{ij}$ is equivalent to the well known scalar
Monaghan-Gingold form of $\Pi_{ij}$ described in \cite{Mreview92} and
\cite{Breview90}.  It is only in the two-dimensional cylindrical Noh problem
presented in \S\ref{NohCyl.sec} that the results using the tensor viscosity are
distinct from those using the normal scalar form of the viscosity.  This is
required because otherwise the errors imposed by the standard viscosity dominate
the solution for this problem, masking the effects we seek to study in this
work.

SPH using Eqs.\ (\ref{sphmass.eq}) -- (\ref{tsphenergy.eq}) has a number of
useful properties.  Conservation of total mass is ensured since the mass is
always simply the sum of the masses of the nodes.  Because
Eq.\ (\ref{tsphmom.eq}) is symmetric for each interacting pair of nodes $i$ and
$j$, the total linear momentum is guaranteed to be conserved because pair-wise
forces are always equal and opposite.  So long as the pair-wise forces are also
radially aligned between points, conservation of angular momentum is also
ensured.  However, if non-radial forces are used (such as is the case with the
tensor viscosity, ASPH \cite{asph98}, or in the presence of material strength),
then the pair-wise forces will no longer be radially aligned (though they remain
equal and opposite), in which case angular momentum is no longer perfectly
preserved.  The energy equation (\ref{tsphenergy.eq}) is intentionally not
symmetrized in the same manner as the momentum equation -- unlike the momentum,
symmetrizing the energy equation does not enforce any particular physical
principles like conservation of momentum.  The nonsymmetrized form of
Eq.\ (\ref{tsphenergy.eq}) has the nice property that it avoids unphysically
cooling nodes to negative temperatures, which can occur if we write the
symmetrized form equivalent to the momentum term.  This issue is discussed in
more detail in \S \ref{fijstandard.sec} below -- also see the discussion of the
Taylor-Sedov blastwave problem in \cite{EntropySPH02}.

We also note that we use a variable smoothing scale $h_i$ associated with each
node.  The algorithm used to update the smoothing scale is substantially
equivalent to that outlined in \cite{Thackaretal00}.  The exact choices for
evolving $h$ do not have a strong impact on the results discussed below -- we
will describe the details of our algorithm for updating the smoothing scale in
future work.

\section{The SPH compatible energy discretization}
\label{compatible.sec}
The central idea of a compatible discretization of a set of physical laws
is that the discrete operations used in numerical computations should exactly
reproduce principles that are enforced in the continuous equations which are
being discretized.  Numerical operations that meet this requirement are said to
be compatible with their continuum counterparts.  In \cite{CompatibleHydro98}
the authors use this principle applied to the Lagrangian fluid dynamics
equations to derive a total energy conserving form of staggered grid Lagrangian
hydrodynamics.  We employ similar reasoning here to derive a total energy
conserving form of SPH -- in this case our job is simplified because all our
physical properties share the same centering, coexisting on the SPH nodes.

We begin by writing down the total energy of the discretized system (ignoring
any external sources or sinks of energy) as
\beq
  E = \sum_i m_i \lp \frac{1}{2} v_i^2 + u_i \rp.
\eeq
We can express the total energy change across a discrete timestep 
(denoting the beginning of timestep values by superscript 0 and end of
timestep values by superscript 1) as
\beq
  \label{DeltaE.eq}
  E^1 - E^0 = \sum_i m_i \left[ \frac{1}{2} \lp v_i^1 \rp^2 + u_i^1 -
                                \frac{1}{2} \lp v_i^0 \rp^2 - u_i^0 \right].
\eeq
Total energy conservation is enforced by setting $E^1 - E^0 = 0$; we can use
$(v_i^\alpha)^1 = (v_i^\alpha)^0 + (a_i^\alpha)^0 \Delta t$ to rewrite
Eq.\ (\ref{DeltaE.eq}) (after some simple algebra) as
\beqa
  0 &=& \sum_i m_i \left[ 
                   \lp (v_i^\alpha)^0 + \frac{1}{2} (a_i^\alpha)^0 \Delta t \rp
                   (a_i^\alpha)^0 \Delta t + u_i^1 - u_i^0 
                   \right] \\
    \label{TotalEBalance.eq}
    &=& \sum_i m_i \left[
                   (v_i^\alpha)^{1/2} (a_i^\alpha)^0 \Delta t + \Delta u_i
                   \right]
\eeqa
where $\Delta t$ is the timestep, $a_i^\alpha$ is the total acceleration on node
$i$, $(v_i^\alpha)^{1/2} = (v_i^\alpha)^0 + (a_i^\alpha)^0 \Delta t/2$ is the
half-timestep velocity, and $\Delta u_i = u_i^1 - u_i^0$ is the specific thermal
energy change.  There are a number of possibilities we could choose for how to
construct $\Delta u_i$ such that Eq.\ (\ref{TotalEBalance.eq}) is met.  One natural
approach is to consider the pair-wise work contribution between any interacting
pair of nodes $i$ and $j$.  We can express the desired total thermal energy
change of the system in terms of the pair-wise interactions as
\beqa
  \DE &=& \sum_i m_i \Delta u_i 
       =  -\sum_i m_i (v_i^\alpha)^{1/2} (a_i^\alpha)^0 \Delta t \nonumber \\
      &=& -\sum_i m_i (v_i^\alpha)^{1/2} \lp \sum_j (a_{ij})^0 \rp \Delta t, \nonumber
\eeqa
where $a_{ij}$ represents the pair-wise contribution to the acceleration of node
$i$ due to node $j$.  The corresponding pair-wise contribution to the total work
is
\beqa
  \label{PairWork.eq}
  \DEij &=& m_i \Delta u_{ij} + m_j \Delta u_{ji} \\
        &=& -\lp m_i (v_i^\alpha)^{1/2} (a_{ij})^0 \Delta t +
                 m_j (v_j^\alpha)^{1/2} (a_{ji})^0 \Delta t \rp \nonumber \\
        &=& m_i \left[ (v_j^\alpha)^{1/2} - (v_i^\alpha)^{1/2} \right]
            (a_{ij})^0 \Delta t, \nonumber
\eeqa
where $\Delta u_{ij}$ represents the specific thermal energy change of node $i$
due to its interaction with node $j$.  Note that in Eq.\ (\ref{PairWork.eq}) we
have explicitly used the fact that pair-wise forces are anti-symmetric ($m_i
a_{ij} = -m_j a_{ji}$), guaranteed by the symmetrization of Eq.\ (\ref{tsphmom.eq}).
This is not a required property to derive our desired compatible energy
equation, however -- it simply removes the necessity of referring to both
$a_{ij}$ and $a_{ji}$ in the equation for node $i$.
Since Eq.\ (\ref{PairWork.eq}) represents the total pair-wise work due to the
interaction of nodes $i$ and $j$, we have the freedom to distribute this work
between these two nodes arbitrarily
\beq
  \label{uij.eq}
  \Delta u_{ij} = f_{ij} \frac{\DEij}{m_i} 
                = f_{ij} \left[ (v_j^\alpha)^{1/2} - (v_i^\alpha)^{1/2} \right]
                  (a_{ij}^\alpha)^0 \Delta t,
\eeq
and exact conservation of the energy is guaranteed so long as $f_{ij} + f_{ji} =
1$.  There are a number of functional forms we could choose for $f_{ij}$ that
meet this constraint; in \S\ref{fijstandard.sec} -- \S\ref{fij.sec} below we consider
several of these forms and describe the choice used in this paper.

The total change in the specific thermal energy for node $i$ is given by
summing over the pair-wise contributions in Eq.\ (\ref{uij.eq}), 
\beqa
  \label{Compui.eq}
  \Delta u_i = \sum_j \Delta u_{ij}
             = \sum_j f_{ij}  \left[ (v_j^\alpha)^{1/2} - (v_i^\alpha)^{1/2} \right]
                      (a_{ij}^\alpha)^0 \Delta t,
\eeqa
and the end of step specific thermal energy is $u_i^1 = u_i^0 + \Delta u_i$.  At
first glance Eq.\ (\ref{Compui.eq}) appears to imply that we are advancing the
specific thermal energy first-order in time, but in fact the issue is a bit more
subtle than that.  First of all, since the velocity we are using is actually
centered at the half-step, this can be viewed as at least partially a
second-order time advancement.  Additionally, in order to maintain compatibility
in a multi-stage time integration scheme we must update the energy according the
half-step velocity for each stage of the integration.  In higher-order (greater
than second-order) time integrators this implies that we would need to compute
an effective time centered velocity and time level zero accelerations at
intermediate stages.  In the end we believe the most consistent way to view the
time advancement of $u_i$ in this formalism is that the specific energy is being
updated at the same order as the velocity, since what we are doing is exactly
accounting for the work done by the accelerations in updating the velocity.  The
total energy is not being ``advanced'' at all since it is guaranteed not to be
changing, whereas in the standard scheme represented by Eq.\
(\ref{tsphenergy.eq}) it is meaningful to refer to the order of the energy
advancement: i.e., the total energy will remain constant to some order of
accuracy.

The compatible SPH discretization therefore uses the standard forms for the
update of the mass and momentum (Eqs.\ \ref{sphmass.eq} and
\ref{tsphmom.eq}) and updates the specific thermal energy according to
Eq.\ (\ref{Compui.eq}), resulting in exact energy conservation to machine
roundoff.  It is important to note, however, that the compatibly differenced form
implies some computational penalties as compared with the standard approach due
to the requirement that the half-step velocity difference $(v_j^\alpha)^{1/2} -
(v_i^\alpha)^{1/2}$ is dotted with the time level 0 pair-wise accelerations
$(a_{ij}^\alpha)^0$.  It is immediately obvious that we must make two passes
over the SPH nodes and their neighbors: once to sum up the accelerations, and a
second time to dot the predicted half-step velocities with the pair-wise
accelerations.  We must also choose to either store the pair-wise accelerations
during the first pass for use in the energy update of the second, or recompute
them during the second pass.  In our own implementation the requirement to make
two passes over the nodes and their neighbors does not impose a large penalty,
largely because we compute and store the set of neighbors for each node at the
beginning of each time step, obviating the necessity of finding those neighbors
on each pass.  We have chosen to store the pair-wise accelerations
$(a_{ij}^\alpha)^0$, paying the memory in order to optimize CPU time.  For this
reason we see very little impact on CPU time in using the compatible
discretization: for example, the 2-D Noh problem discussed in \S
\ref{NohCyl.sec} shows $\sim$ 3\% increase in the cycle time for the compatible
discretization compared with the standard method.  However, we do pay a
substantial cost in memory to store the pair-wise accelerations.  In our
experience SPH is typically CPU bound rather than memory bound, which is why we
have made these choices.  However, on some computer architectures memory
constraints could well dominate and it may be more beneficial to recompute the
pair-wise accelerations rather than store them.

\subsection{Choosing \fij}
\subsubsection{Comparison with the standard work distribution.}
\label{fijstandard.sec}
One natural choice we could make for selecting the pair-wise work distribution
$\fij$ is an equipartition of the pair-wise work, $f_{ij} = 1/2$.  Plugging this
choice into Eq.\ (\ref{uij.eq}) and substituting in the pair-wise acceleration
from Eq.\ (\ref{tsphmom.eq}) yields
\beq
  \label{duijsym.eq}
  \Delta u_{ij} = \frac{1}{2} \left[
                  m_j \lp \frac{P_i}{\rho_i^2} + \frac{P_j}{\rho_j^2} \rp
                  (v_{ij}^\alpha)^{1/2}
                  \partial_\alpha \Wij +
                  \Pi^{\alpha \beta}_{ij}
                  (v_{ij}^\beta)^{1/2}
                  \partial_\alpha \Wij \right] \Delta t.
\eeq
This is nearly identical to the pair-wise work that results from the fully
symmetrized form of the standard energy equation
\beq
  \frac{Du_i}{Dt} = \frac{1}{2} \sum_j m_j \left[
                    \lp \frac{P_i}{\rho_i^2} + \frac{P_j}{\rho_j^2} \rp
                    v_{ij}^\alpha \partial_\alpha \Wij +
	            \Pi^{\alpha \beta}_{ij}
                    v_{ij}^\beta \partial_\alpha \Wij
                    \right],
\eeq
with the minor exception that the velocity difference is performed at the
half-timestep in the compatible formalism.  However, as pointed out previously
this approach has the weakness that it can unphysically over-cool nodes.  This
is easily seen if we consider a very hot node interacting with a node at zero
specific thermal energy when the net work is cooling.  According to
Eq.\ (\ref{duijsym.eq}) the cooling energy change will be equipartitioned between
the nodes, and the node with zero initial energy will be forced to negative
thermal energy.  For this reason we do not use this definition for $f_{ij}$.

Another possibility results if we distinguish between the pressure and viscous
contributions to the work, choosing
\beq
  f_{ij}^{P dV} = \frac{P_i/\rho_i^2}{P_i/\rho_i^2 + P_j/\rho_j^2}
\eeq
for the acceleration due to the pressure, while maintaining the equipartitioning
definition $f_{ij}^{\Pi} = 1/2$ for the artificial viscous term.  Substituting
these definitions into Eqs.\ \ref{uij.eq} and \ref{tsphmom.eq} we now find
\beq
  \Delta u_{ij} = m_j \left[ \frac{P_i}{\rho_i^2} 
                  (v_{ij}^\alpha)^{1/2}
                  \partial_\alpha \Wij +
                  \frac{1}{2} \Pi^{\alpha \beta}_{ij}
                  (v_{ij}^\beta)^{1/2}
                  \partial_\alpha \Wij \right] \Delta t.
\eeq
This form reproduces the pair-wise change from Eq.\ (\ref{tsphenergy.eq}), again with
the exception that the velocity difference is performed at the half-step.  This
choice is one reasonable possibility for $f_{ij}$, though in the following
sections we will seek to improve upon this.

\subsubsection{A smoothly variation dimensioning approach.}
Another approach we can adopt in distributing the work between pairs of nodes is
to try and reduce the variation in the specific thermal energy, in an attempt to
reduce the spurious introduction of new extrema.  In other words, we can choose
a form for $\fij$ such that for $\DEij < 0$ we preferentially cool the hotter
of $i$ and $j$, whereas if $\DEij > 0$ we prefer to heat the cooler of the pair.

\begin{figure}[htb]
\turnbox{90}{$\fijsm$}
\parbox[h]{0.8\textwidth}{
\begin{center}
\includegraphics*[width=0.8\textwidth]{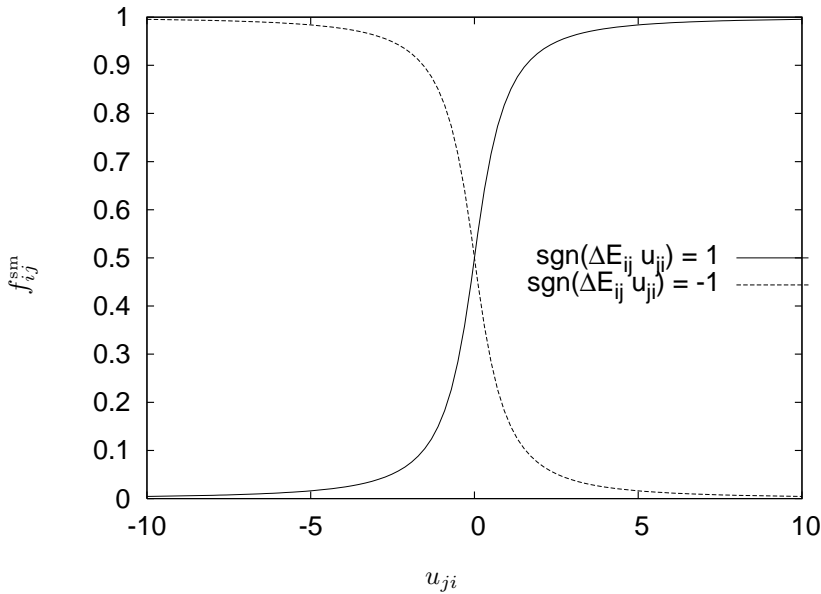} \\
$u_{ji}$
\end{center}
}
\caption{The weighting function $\fijsm$ as defined in Eq.\ (\ref{fijsmooth.eq}).}
\label{fijsmooth.fig}
\end{figure}

One simple analytic representation for $\fij$ that meets these criteria is
\beq
  \label{fijsmooth.eq}
  \fijsm = \frac{1}{2} \left[ 1 +
           \frac{u_{ji} \; \sgn{\DEij}}{|u_{ji}| + (1 + |u_{ji}|)^{-1}}
           \right],
\eeq
where $u_{ji} = u_j - u_i$ and $\sgn{x}$ is the usual sign function given as 
\beq
  \sgn{x} = \left\{
  \begin{array}{r@{\quad:\quad}l}
   -1 & x < 0, \nonumber \\
    0 & x = 0, \\
    1 & x > 0. \nonumber \\
  \end{array} \right.
\eeq
Figure \ref{fijsmooth.fig} shows $\fijsm$ for a range of $u_{ji}$
around 0.  This (admittedly arbitrary) choice for $f_{ij}$ meets the criteria
outlined above: it drives the temperatures of nodes $i$ and $j$ together
(discouraging new extrema), goes to the limit $\fij = 1/2$ if either $u_{ji} =
0$ or $\DEij = 0$ implying an equipartitioning of the work, and meets the
condition $f_{ij} + f_{ji} = 1$ required to maintain energy conservation.  This
definition for $\fij$ works quite well in practice.

\subsubsection{A strictly monotonic variation diminishing approach.}
A potential shortcoming of Eq.\ (\ref{fijsmooth.eq}) is that it can still create
new overall extrema, because it will move both $u_i$ and $u_j$ in response to
work done by the accelerations.  This suggests another possibility: we could define
a strictly monotonic form for $\fij$ which deposits all of the work on node
$i$ or $j$ (i.e., all the heating on the cooler node or all the cooling on the
hotter) so long as the resulting specific energy is bounded by the original
range: $u_i^1 \in [u_i^0, u_j^0]$ and $u_j^1 \in [u_i^0, u_j^0]$.  If the
pair-wise work will result in violating these bounds (implying the
heating/cooling is larger than the original specific energy discrepancy between
$i$ and $j$), then we deposit the work such that the final specific energies are
identical: $u_i^1 = u_j^1 = u^1$.  This idea leads to the following definition
for $\fijmono$,
\beq
  A = \frac{\DEij}{u_{ji}}, \; \;
  B = \left\{
      \begin{array}{r@{\quad: \quad}l}
      A/m_i & A \ge 0, \\
      A/m_j & A < 0, \nonumber
      \end{array} \right.
\eeq
\beq
  \label{fijmono.eq}
  \fijmono = \left\{
             \begin{array}{l@{\quad: \quad}l}
             \max(0, \; \sgn{B}) & |B| \le 1, \\
             \frac{m_i}{\DEij} 
             \left( \frac{\DEij + m_i u_i^0 + m_j u_j^0}{m_i + m_j} - u_i^0
             \right) & |B| > 1. \nonumber \\
             \end{array} \right.
\eeq
Simulations employing $\fijmono$ show improved monotonicity in many problems
(particularly near discontinuities), but in general do not result in the same degree
of accuracy as the smoother definition of $\fijsm$ in Eq.\ (\ref{fijsmooth.eq}).

\subsubsection{A hybrid definition for $\fij$}
\label{fij.sec}
Based on our experience experimenting with these various choices for $\fij$, we
have settled on a hybrid definition that combines $\fijsm$
(Eq.\ \ref{fijsmooth.eq}) and $\fijmono$ (Eq.\ \ref{fijmono.eq}), defined as
\beq
  \chi = \frac{|u_j - u_i|}{|u_i| + |u_j| + \zeta},
\eeq
\beq
  \label{fij.eq}
  \fij = \chi \fijmono + (1 - \chi) \fijsm,
\eeq
where the $\zeta$ term is a small number to avoid division by zero.  Clearly the
$\chi$ weighting transitions smoothly from the smooth case $\fijsm$ so long as
$u_i$ and $u_j$ are relatively similar (compared with their average values) to
the monotonic form $\fijmono$ as the difference between $u_i$ and $u_j$ widens.
If either $u_i$ or $u_j$ is zero (or they have opposite signs), this definition
recovers the monotonic form.  We have found this hybrid to be quite robust in
practice, and this is the form used for all the examples in this paper.

\section{Tests}
\label{tests.sec}
In this section we compare the results of applying the standard SPH
discretization (Eqs.\ \ref{sphmass.eq} -- \ref{tsphenergy.eq}) vs. the compatibly
differenced form (Eqs.\ \ref{sphmass.eq}, \ref{tsphmom.eq}, \& \ref{Compui.eq})
on a variety of standard shock driven test cases.  Each of the tests presented
here has an analytic solution, allowing us to measure the $L_p$ error norms of
the simulated results against these analytic answers.  We will use these error
comparisons to quantitatively compare the accuracy and convergence rates of the
different techniques.  We use the definition of the $p$th error norm
for a set of $N$ values $x_i$ compared with an analytic answer $y_i$ as
\beq
  \label{Lp.eq}
  L_p(x) = \lp \frac{\sum_{i = 1}^{N} |x_i - y_i|^p}{N} \rp^{1/p},
\eeq
which in the limit $p \to \infty$ becomes
\beq
  \label{Lpinf.eq}
  L_\infty(x) = \max(|x_i - y_i|).
\eeq

A natural question that arises when considering energy conserving schemes such
as this is, what about the entropy?  It is well known that algorithms which
achieve energy conservation by advancing the total energy are prone to serious
errors in the entropy evolution, particularly in the presence of strongly
supersonic flows.  We do not expect the compatible discretization to suffer this
weakness, as we are still evolving the specific internal energy via
Eq.\ (\ref{Compui.eq}) -- we are simply making a more accurate accounting of the
discrete work done due to the discrete momentum relation (Eq.\ \ref{tsphmom.eq}).
In order to directly address this concern in the following tests we will
explicitly examine the evolution of the entropic function
\beq
  A(s) = \frac{P}{\rho^\gamma},
\eeq
as described in \cite{EntropySPH02}.

\subsection{Planar Sod shock tube}
\label{SodPlanar.sec}
\begin{figure}[htb]
\begin{minipage}[b]{\textwidth}
\begin{tabular}[h]{rccc}
 & $\rho$ & $v$ & $A(s)$ \\
\turnbox{90}{\hspace{0.05\textwidth}Standard} &
\subfigure{
\includegraphics*[width=0.35\textwidth]{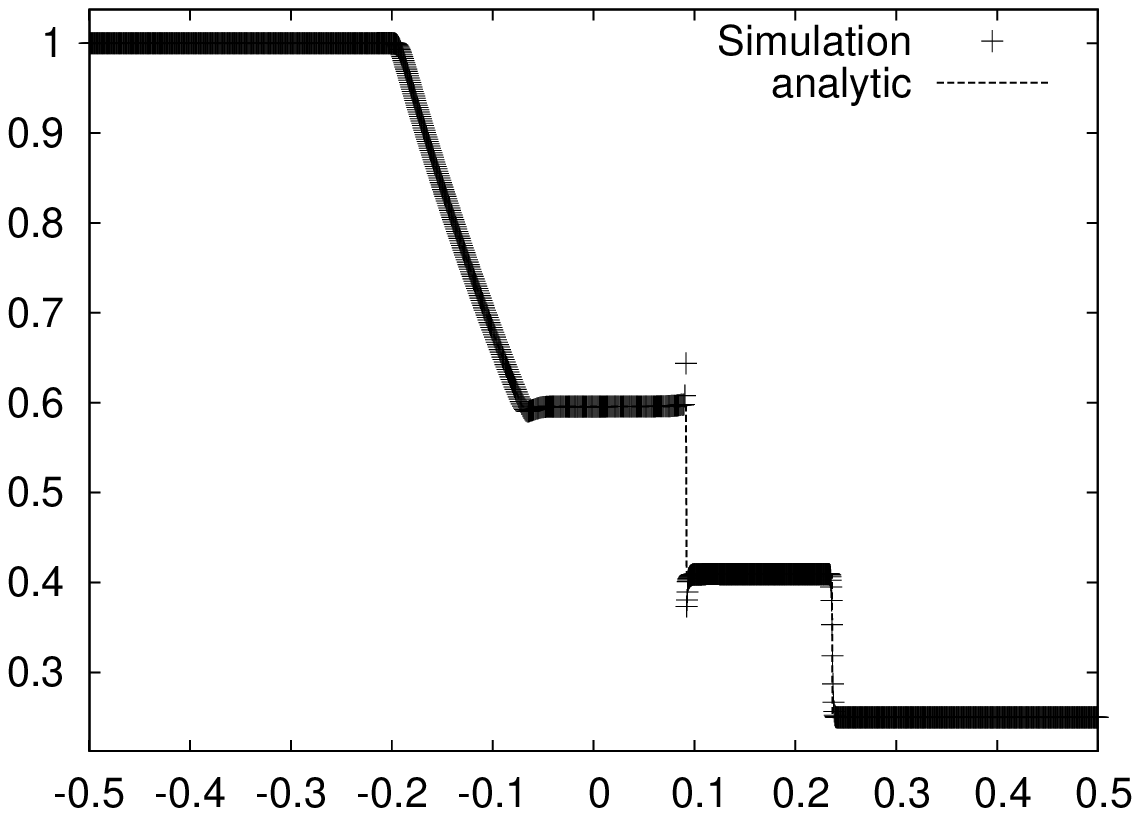}
} &
\subfigure{
\includegraphics*[width=0.35\textwidth]{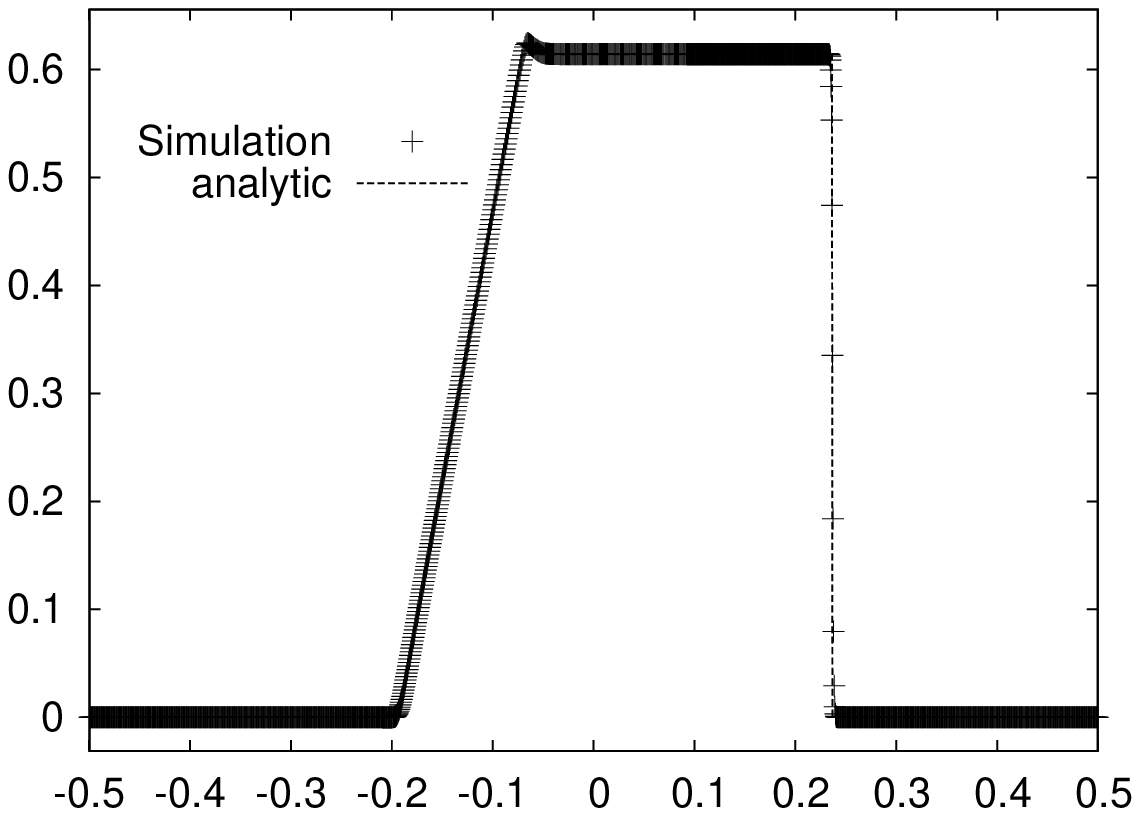}
} &
\subfigure{
\includegraphics*[width=0.35\textwidth]{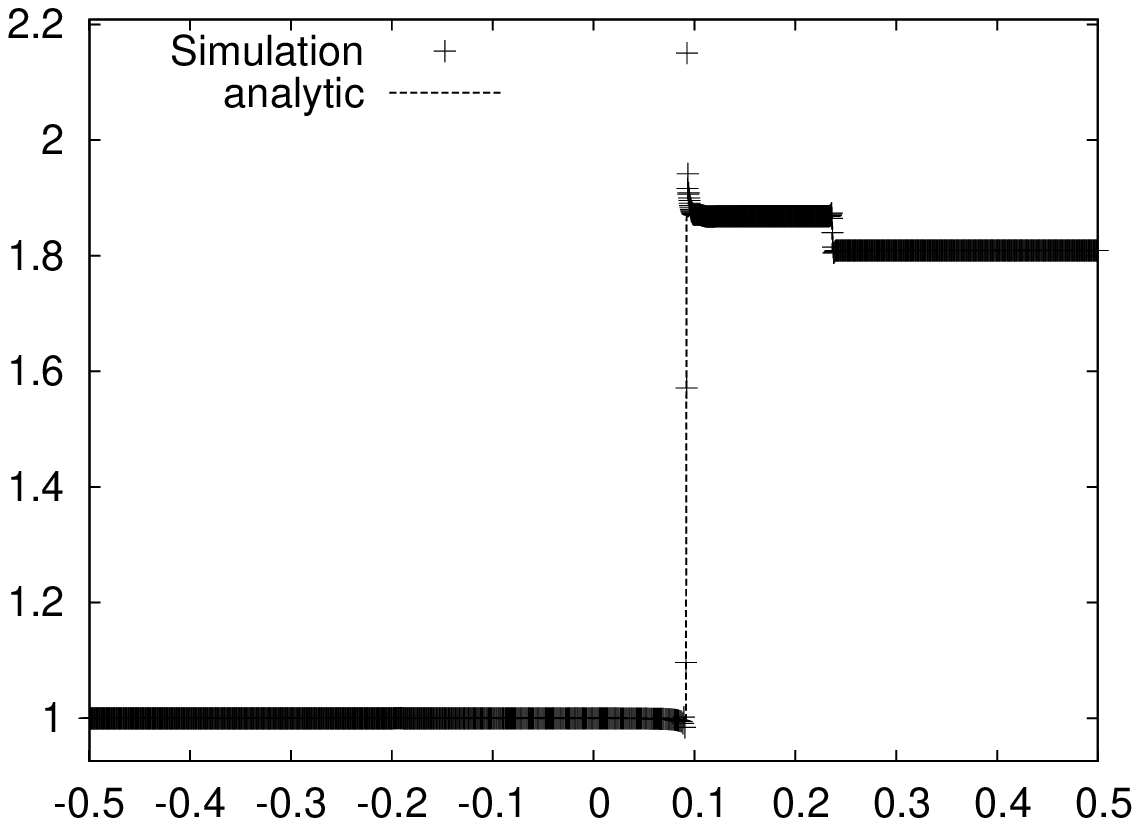}
} \\
\turnbox{90}{\hspace{0.1\textwidth}Compatible} &
\subfigure{
\includegraphics*[width=0.35\textwidth]{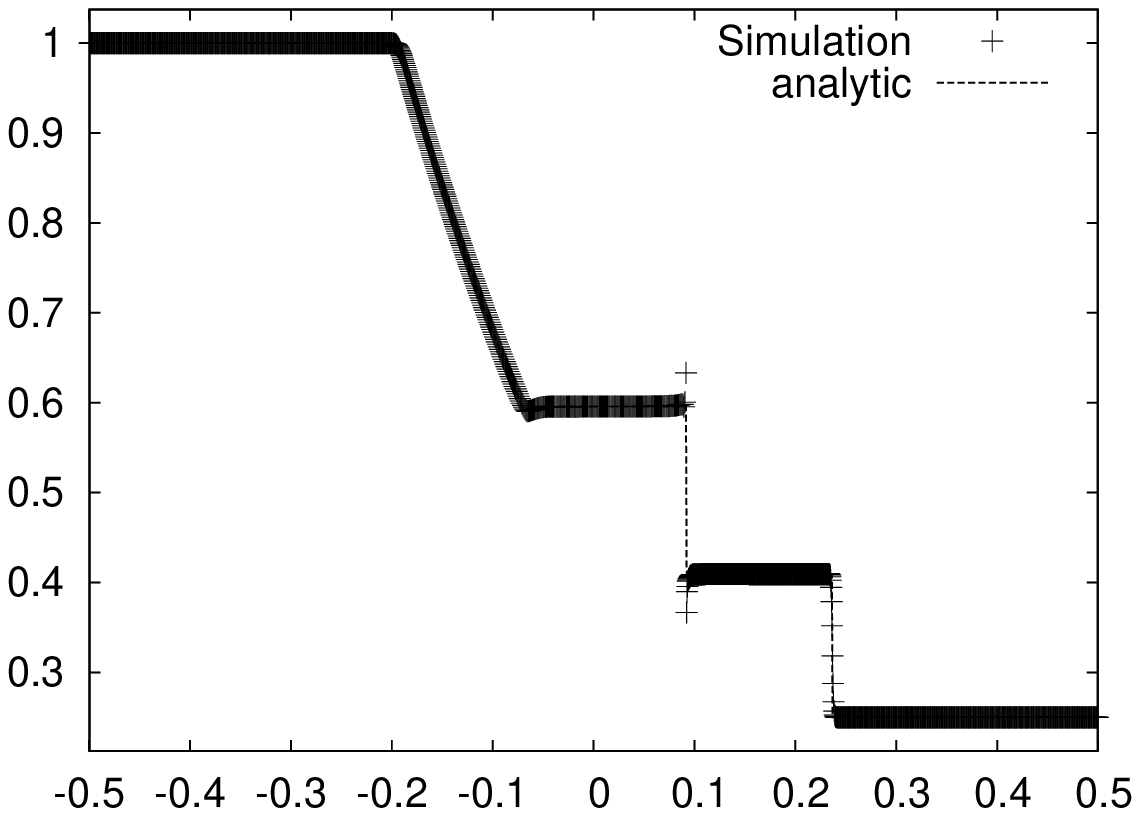}
} &
\subfigure{
\includegraphics*[width=0.35\textwidth]{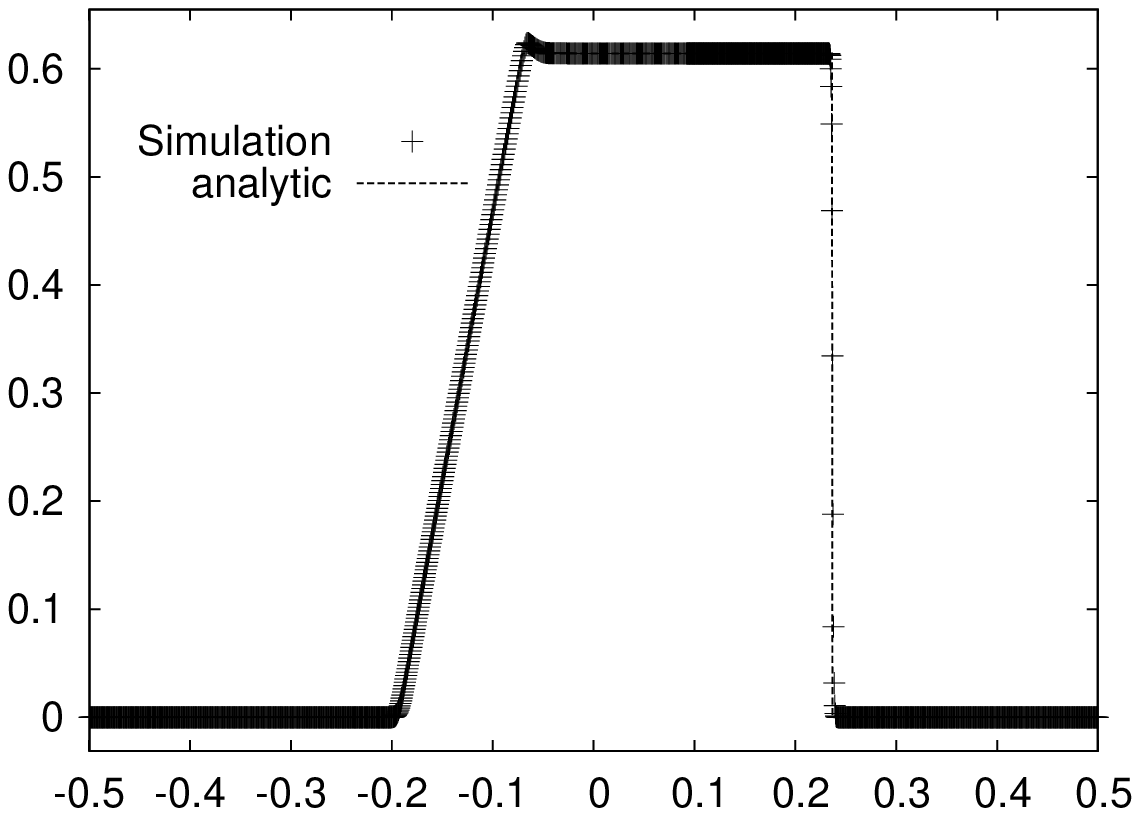}
} &
\subfigure{
\includegraphics*[width=0.35\textwidth]{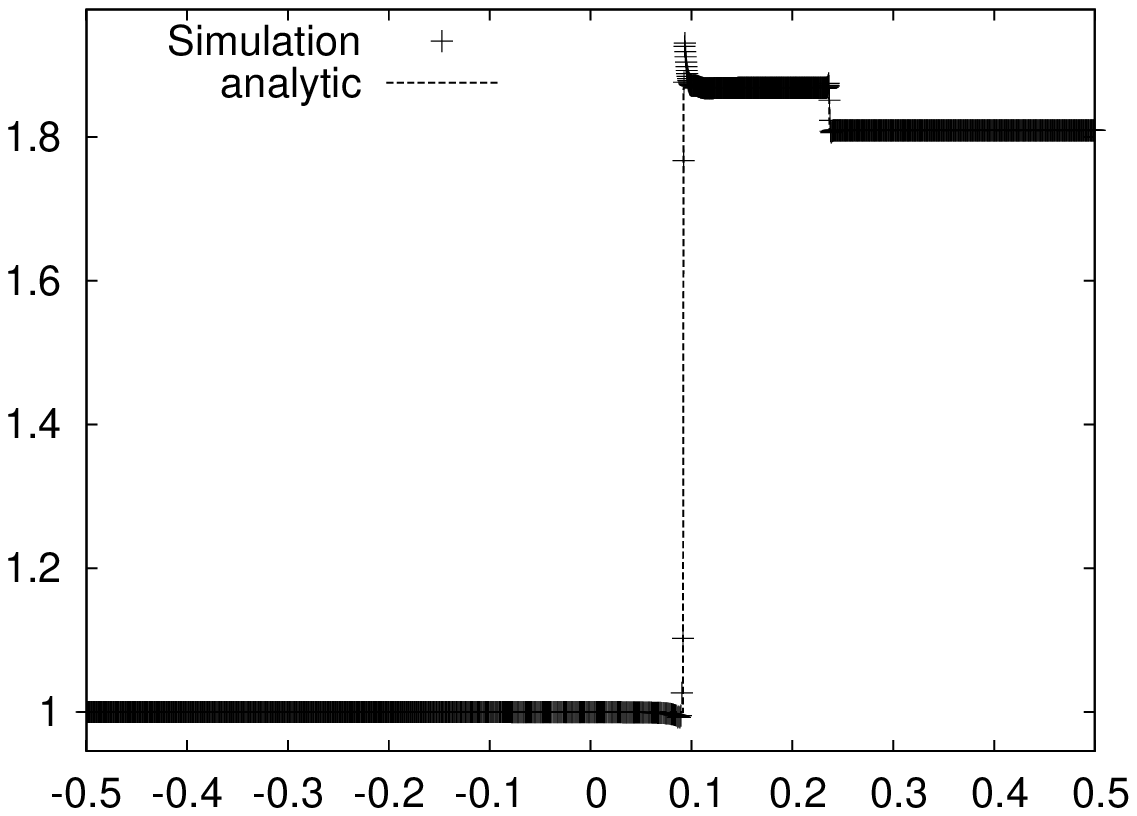}
}
\end{tabular}
\centerline{$x$}
\end{minipage}
\caption{Mass density, velocity, and $A(s)$ profiles for the 800 node simulations
  of the Sod test case, compared with the analytic answer.
}
\label{PlanarSodProfiles.fig}
\end{figure}
We first consider the planar Sod shock tube test case \cite{Sod78}.  This
problem assumes that initially two regions of gas with different densities and
pressures are separated by a membrane which at time $t=0$ is punctured, allowing
the gases to relax into one and other.  This results in a shock propagating into
the low pressure material while a rarefaction propagates into the higher
pressure region.  We consider the standard version most often tested in SPH
\cite{MG83}: we assume a $\gamma = 5/3$ gamma-law gas with initial conditions in
the high pressure region $(\rho, P, v) = (1, 1, 0)$, while the low pressure
region has $(\rho, P, v) = (0.25, 0.1795, 0)$.  This scenario results in a
relatively weak shock.  We initialize this problem with equal numbers of SPH
points in each region, testing cases with total numbers of SPH nodes $N \in$
(100, 200, 400, 800, 1600, 3200, 6400, \& 12800).

Figure \ref{PlanarSodProfiles.fig} shows the profiles of the mass densities,
velocities, and entropic function for 800 node simulation at time $t = 0.15$.
To the eye both the standard and compatible discretizations do well on this
problem: the only evident difference is that the overshoot in $A(s)$ at the
Lagrangian position of the initial discontinuity is improved using the compatible
formalism.

\begin{figure}[htb]
\begin{minipage}[b]{\textwidth}
\begin{tabular}[h]{rccc}
 & $\rho$ & $v$ & $A(s)$ \\
\turnbox{90}{\hspace{0.1\textwidth}Error} &
\subfigure{
\includegraphics*[width=0.38\textwidth]{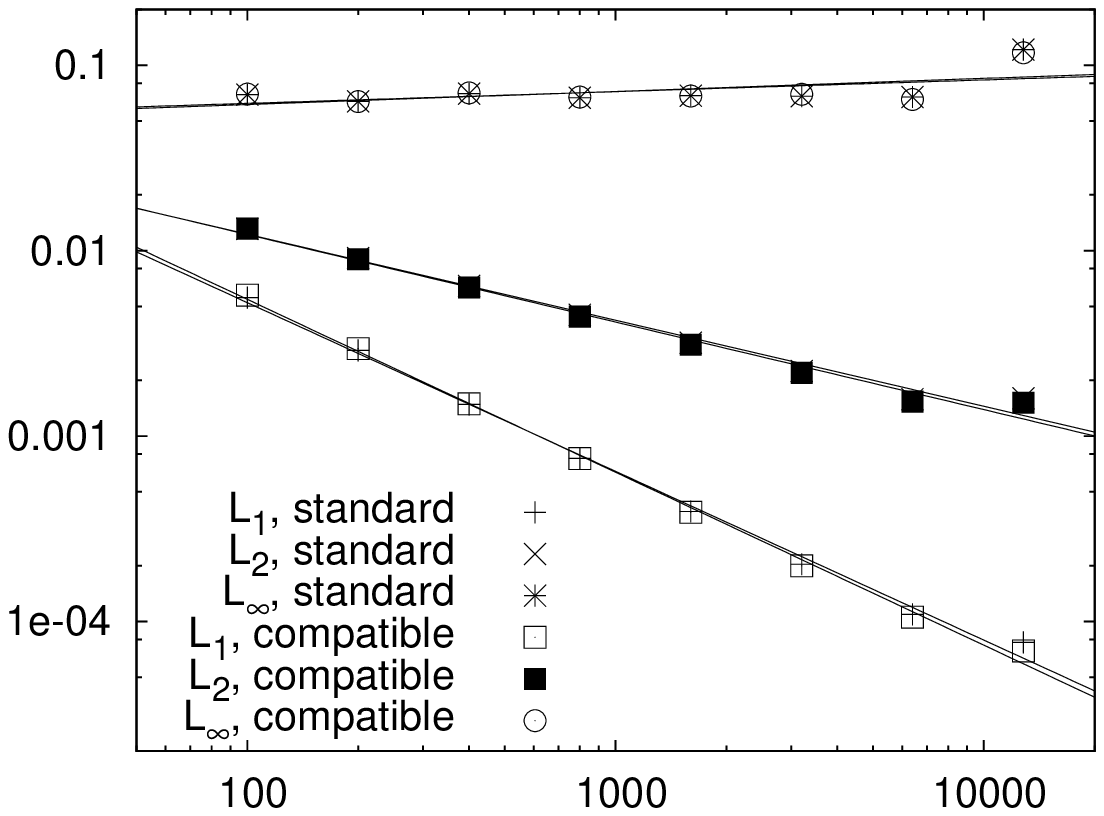}
} & 
\subfigure{
\includegraphics*[width=0.38\textwidth]{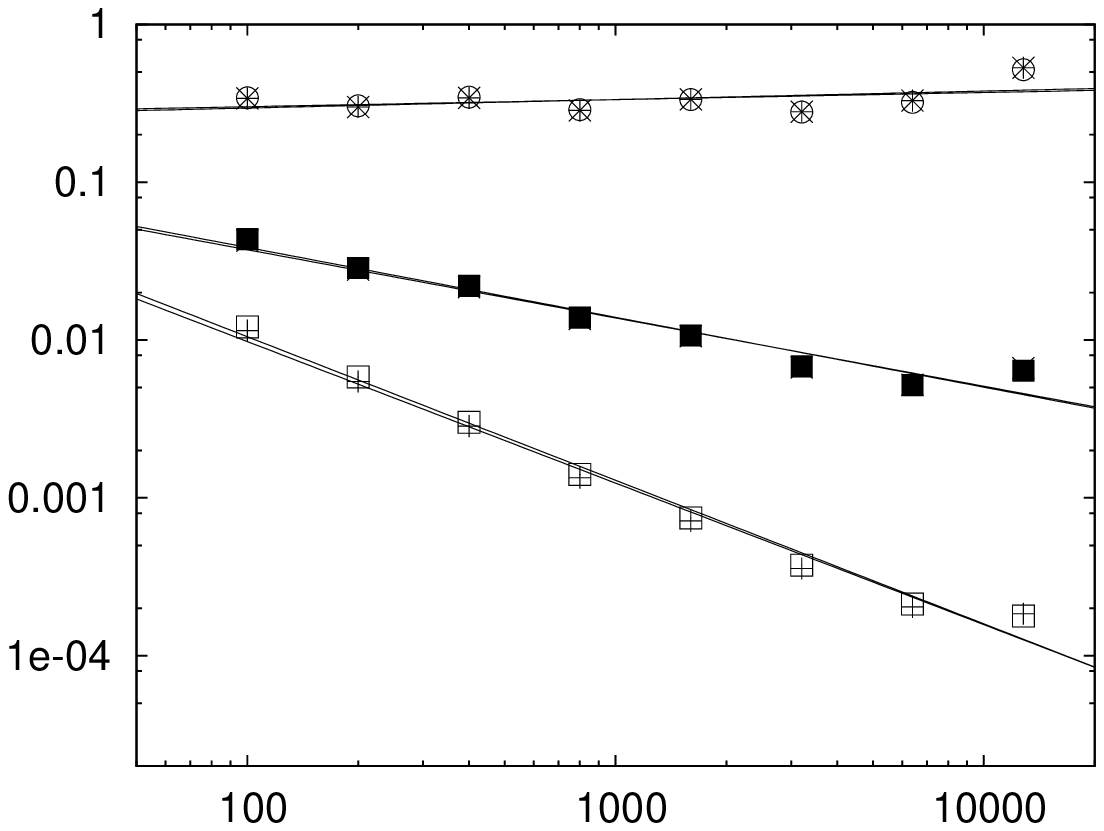}
} & 
\subfigure{
\includegraphics*[width=0.38\textwidth]{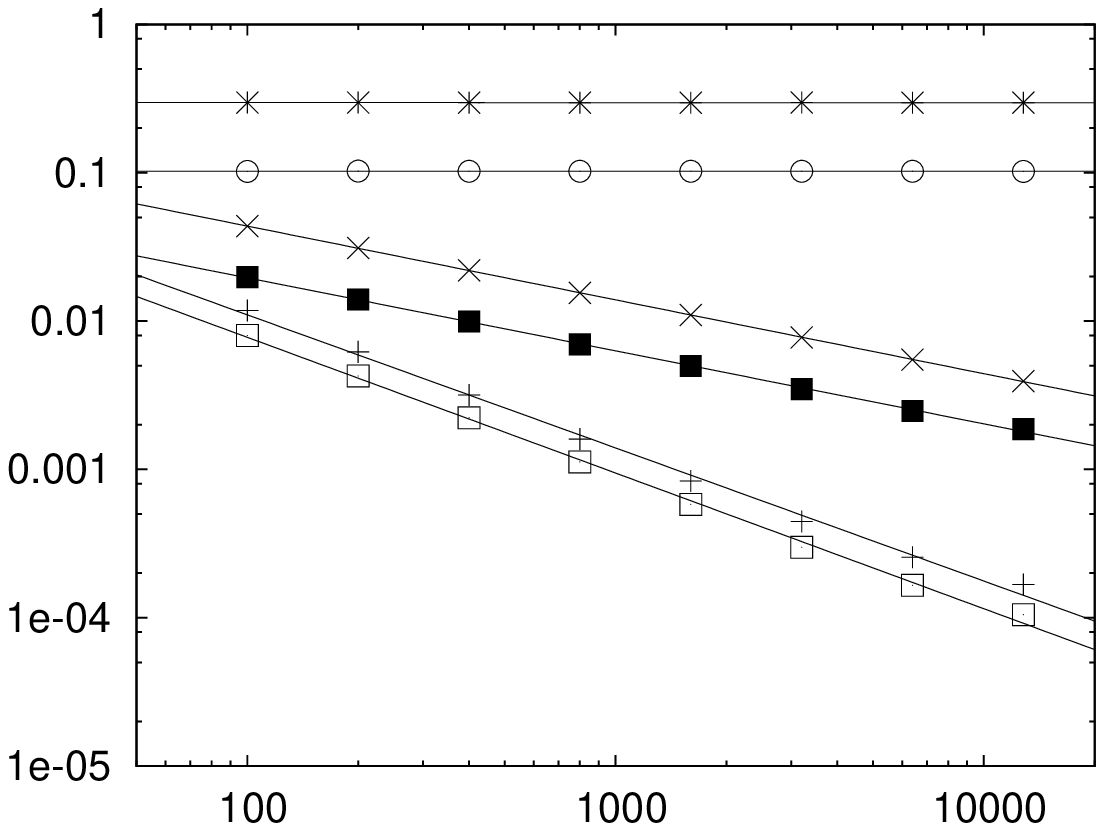}
}
\end{tabular}
\centerline{$N$}
\end{minipage}
\caption{Error estimates ($L_1$, $L_2$, and $L_\infty$, points) and the
corresponding fitted convergence rates (lines) for the mass density,
velocity, and entropic function $A(s)$ in the 1-D planar Sod test case.
}
\label{PlanarSodErrors.fig}
\end{figure}

\begin{table}
\begin{tabular}[htb]{|cr|c|c|c|} \hline
 &  & $L_1$ & $L_2$ & $L_\infty$ \\ \hline
 &  Std. & -0.91 $\pm$ 0.02 & -0.46 $\pm$ 0.02 & 0.07 $\pm$ 0.04 \\
\raisebox{2.5ex}{$\rho$} 
 & Comp. & -0.93 $\pm$ 0.02 & -0.47 $\pm$ 0.02 & 0.06 $\pm$ 0.04 \\ \hline
 &  Std. & -0.89 $\pm$ 0.05 & -0.43 $\pm$ 0.04 & 0.05 $\pm$ 0.04 \\
\raisebox{2.5ex}{$v$} 
 & Comp. & -0.91 $\pm$ 0.04 & -0.44 $\pm$ 0.04   & 0.04 $\pm$ 0.04 \\ \hline
 &  Std. & -0.90 $\pm$ 0.02 & -0.498 $\pm$ 0.001 & -0.0001 $\pm$ 0.0001 \\
\raisebox{2.5ex}{$A(s)$} 
 & Comp. & -0.92 $\pm$ 0.02 & -0.492 $\pm$ 0.005  & -0.0003 $\pm$ 0.0004 \\ \hline
\end{tabular}
\caption{Fitted convergence rates for the mass density, velocity, and $A(s)$ in the
  planar Sod problem, shown $\pm 1 \sigma$.
}
\label{SodConv.tab}
\end{table}
Figure \ref{PlanarSodErrors.fig} shows the $L_1$, $L_2$, and $L_\infty$ error
measurements for the Sod test results as defined in Eqs.\ (\ref{Lp.eq}) \&
(\ref{Lpinf.eq}).  We also plot the fitted convergence rates for these error
measurements, the slopes of which are summarized in Table \ref{SodConv.tab}.
The trend we see where the higher order error norms show poorer convergence is
expected in shock dominated problems such as this.  This is a result of the fact
that we are trying to fit a step function in the physical properties with a
smooth curve: as we increase the resolution of the simulation the percentage of
points involved in the smoothing over the step function shrinks (which is why
the low-order norms improve), but the maximum discrepancy remains essentially
constant, leading to a constant $L_\infty$.  This is precisely the behaviour we
see in Fig. \ref{PlanarSodErrors.fig}.  We can see that errors and associated
convergence rates are comparable in the two methods for the mass density and
velocity, but the errors in $A(s)$ are reduced by using the compatible
discretization.  Both techniques achieve the expected first-order convergence
rate (resulting in a power $m = -1$ for $L_1(n_1)/L_1(n_2) = (n_1/n_2)^m$),
which is the best we can expect since this is a shock dominated problem.

The energy conservation in this problem is excellent.  If we define the energy
drift as 
\beq
  \frac{\Delta E}{E} = \frac{E^{\mbox{\scriptsize final}} - E^{\mbox{\scriptsize initial}}}
                            {E^{\mbox{\scriptsize initial}}},
\eeq
then we find the maximum energy error for the standard case is $\Delta E/E \sim
2 \times 10^{-5}$.  The compatible discretization of course conserves energy
manifestly, so the energy error (in this case at most $\sim 10^{-15}$) is more a
debugging check than a measure of how well the simulation is progressing.

\subsection{Planar Noh shock test}
\label{NohPlanar.sec}
\begin{figure}[htb]
\begin{minipage}[b]{\textwidth}
\begin{tabular}[h]{rccc}
 & $\rho$ & $v$ & $A(s)$ \\
\turnbox{90}{\hspace{0.05\textwidth}Standard} &
\subfigure{
\includegraphics*[width=0.35\textwidth]{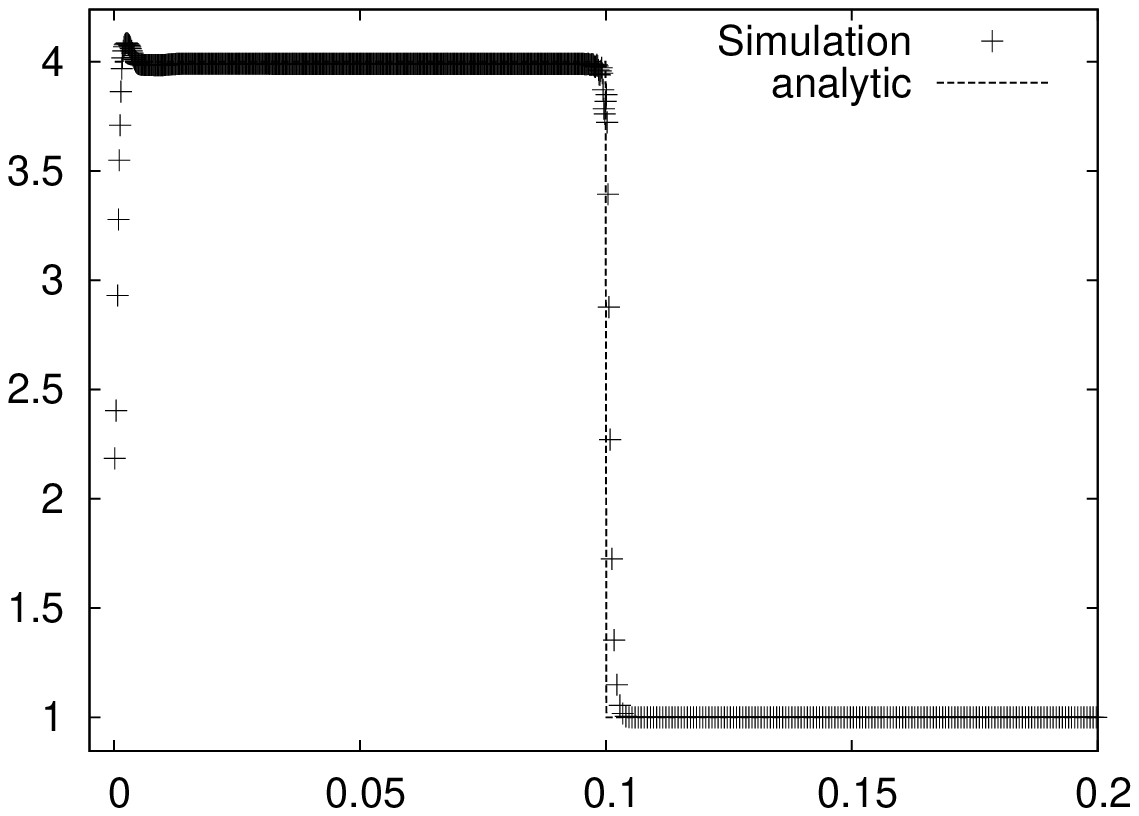}
} &
\subfigure{
\includegraphics*[width=0.35\textwidth]{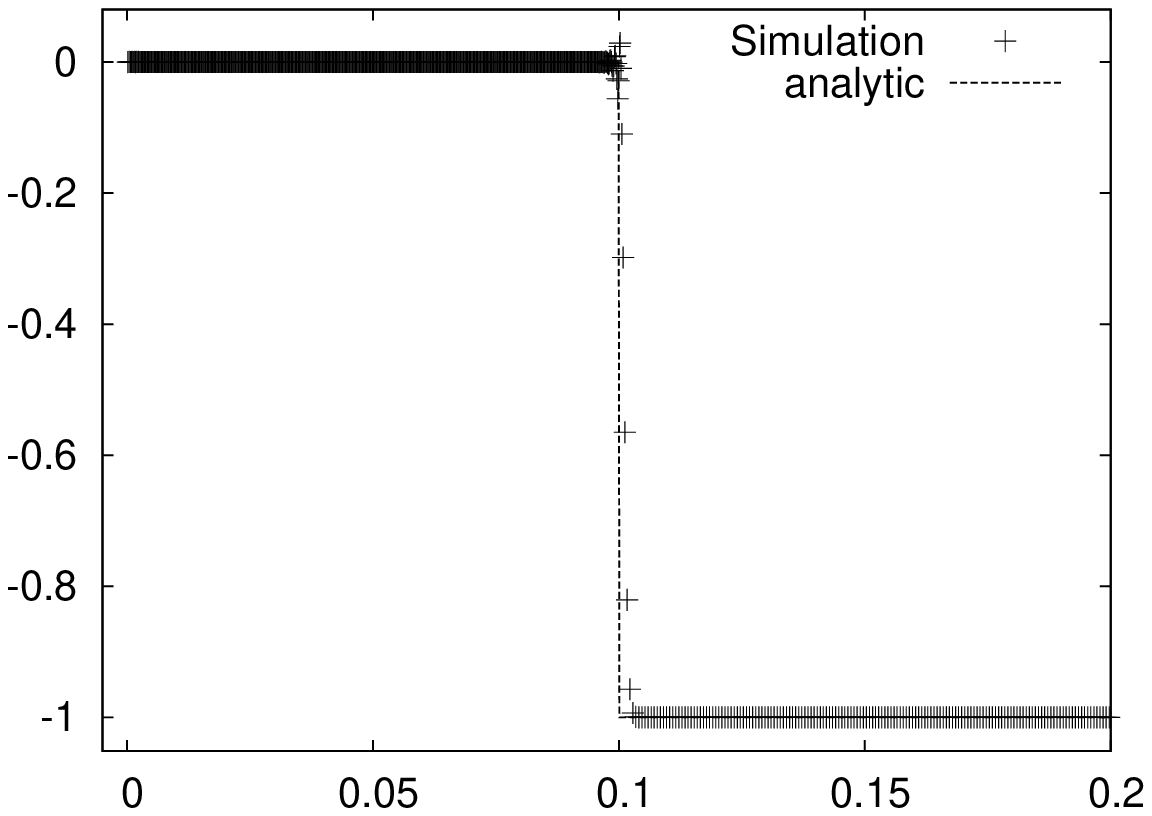}
} &
\subfigure{
\includegraphics*[width=0.35\textwidth]{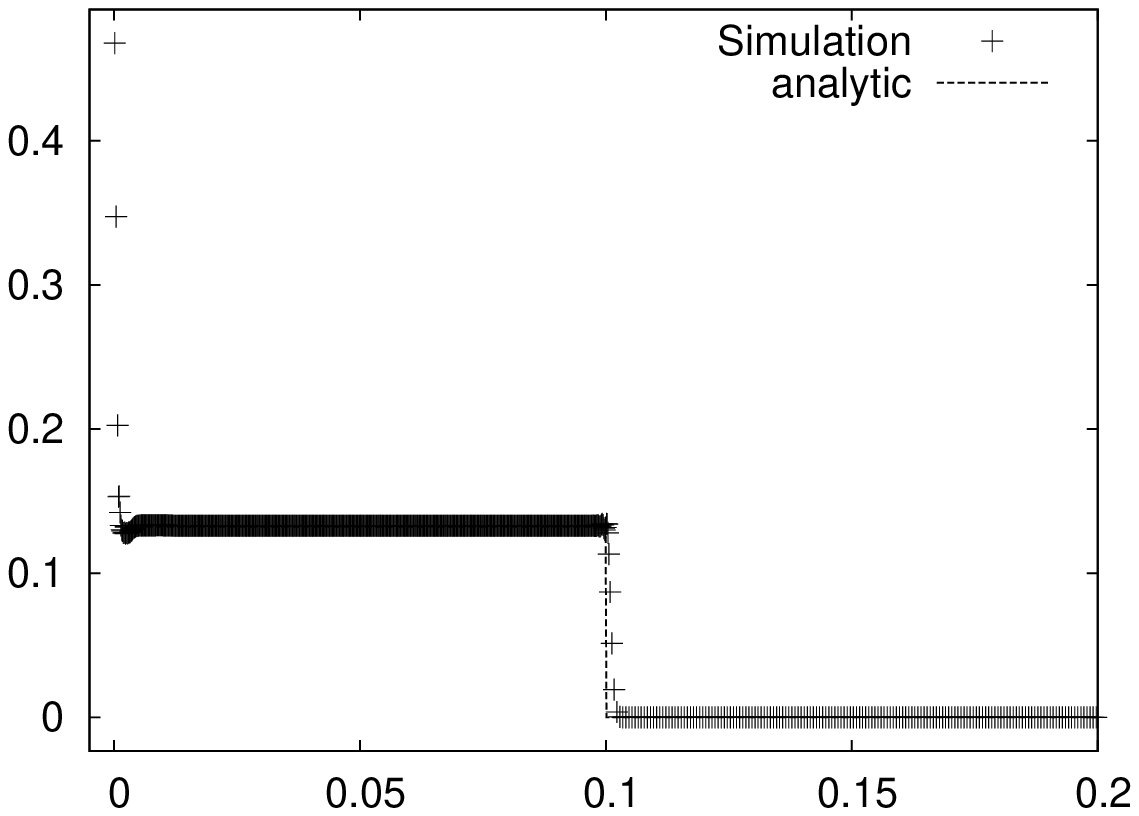}
} \\
\turnbox{90}{\hspace{0.1\textwidth}Compatible} &
\subfigure{
\includegraphics*[width=0.35\textwidth]{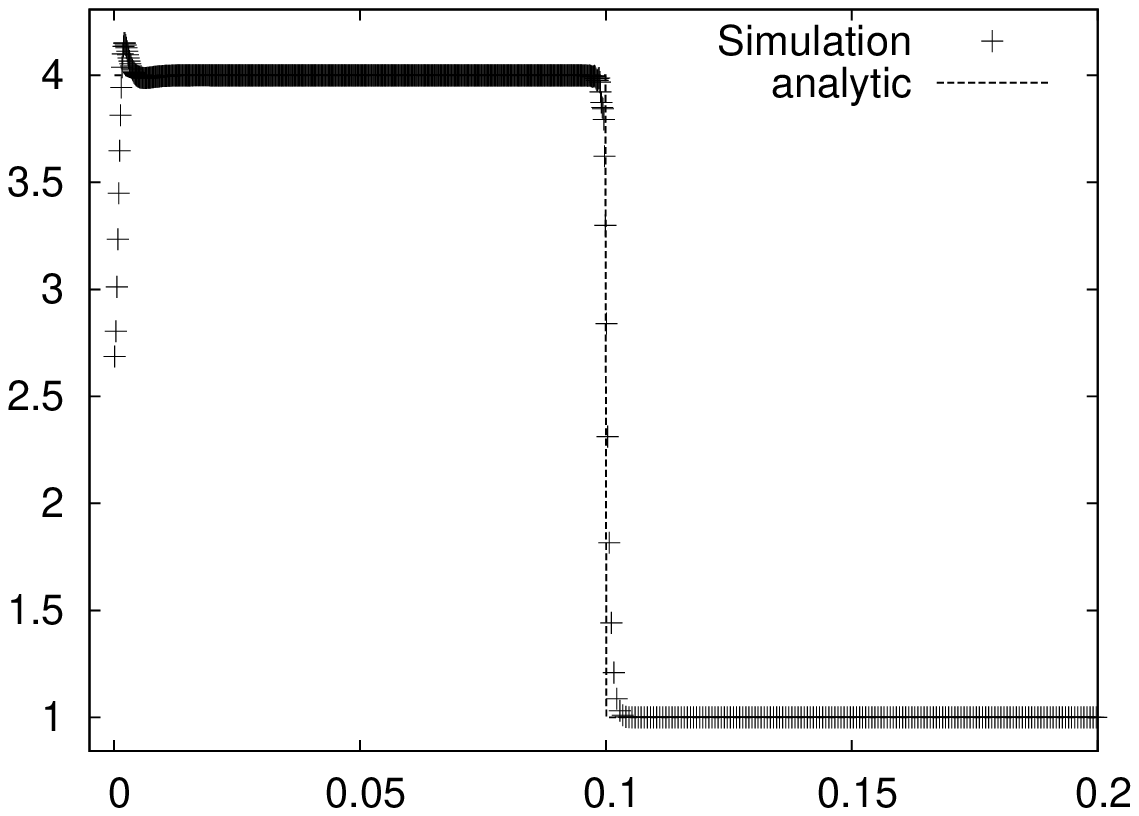}
} &
\subfigure{
\includegraphics*[width=0.35\textwidth]{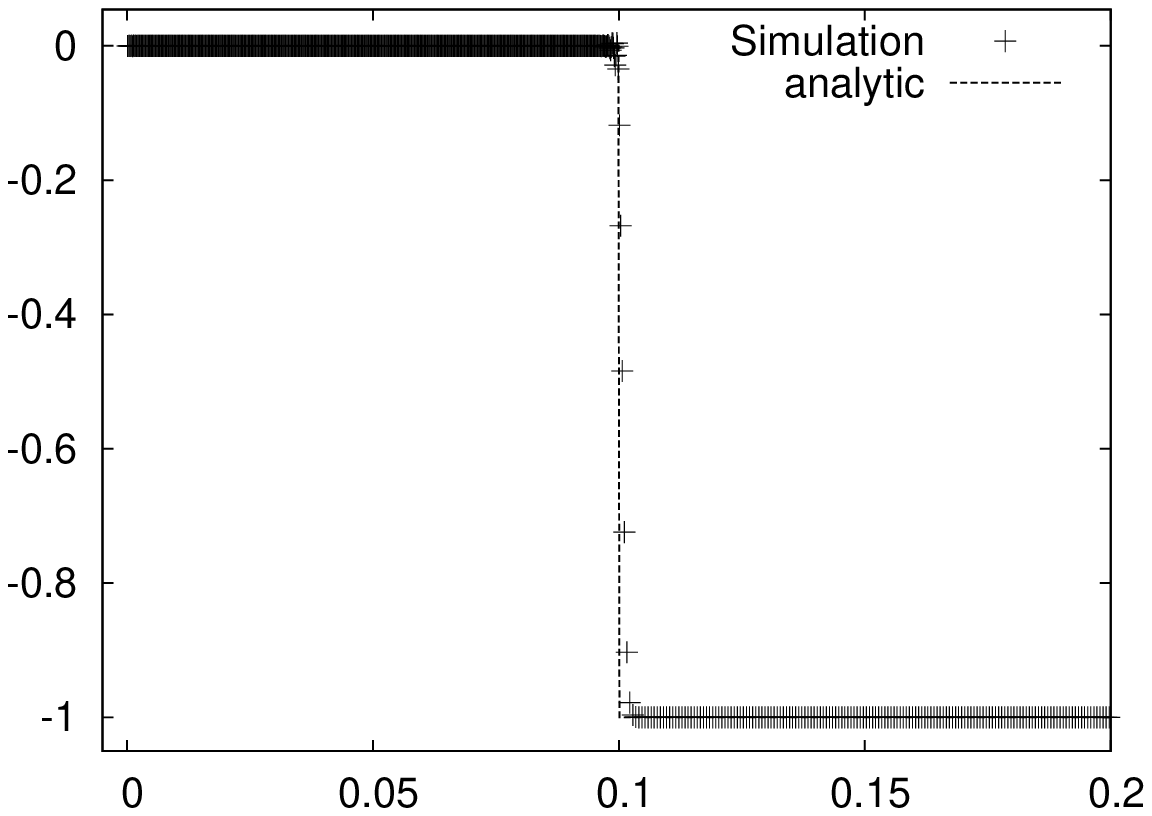}
} &
\subfigure{
\includegraphics*[width=0.35\textwidth]{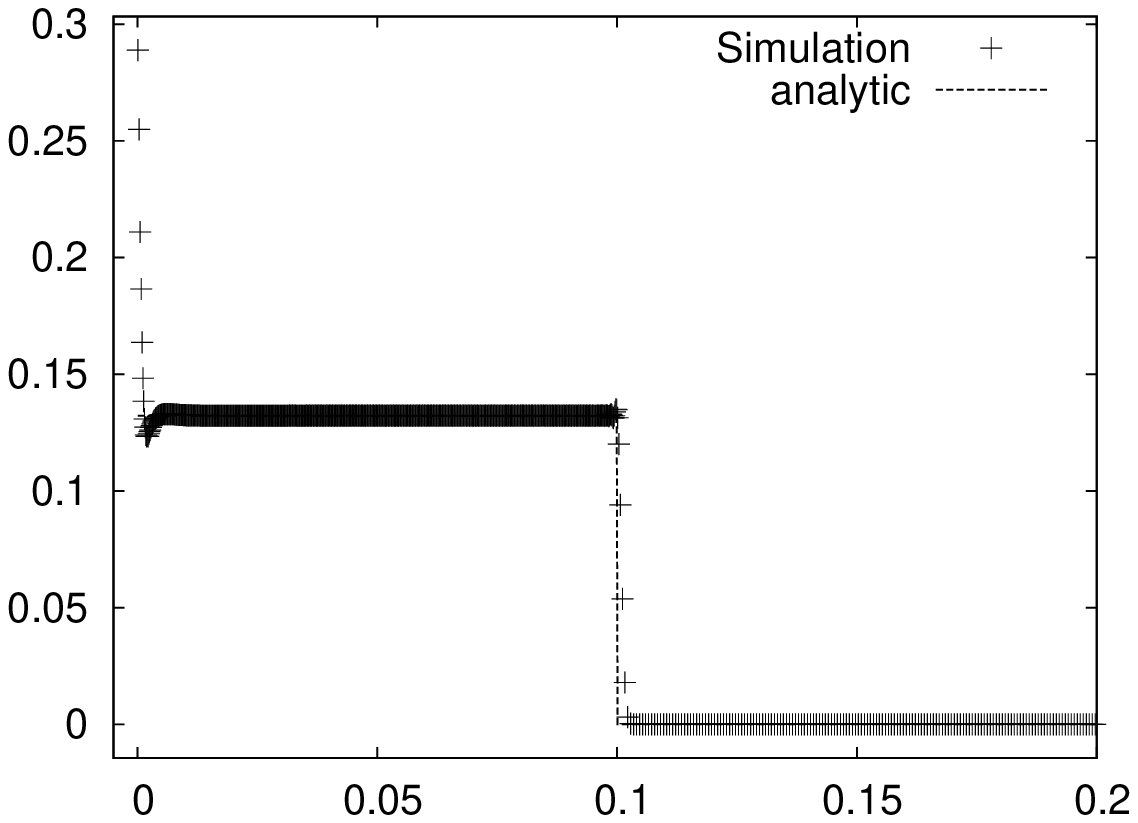}
}
\end{tabular}
\centerline{$x$}
\end{minipage}
\caption{Mass density, velocity, and $A(s)$ profiles for the 800 node simulations of the
  planar Noh shock test case compared with the analytic solution.
}
\label{PlanarNohProfiles.fig}
\end{figure}

\begin{figure}[htb]
\begin{minipage}[b]{\textwidth}
\begin{tabular}[h]{rccc}
 & $\rho$ & $v$ & $A(s)$ \\
\turnbox{90}{\hspace{0.1\textwidth}Error} &
\subfigure{
\includegraphics*[width=0.38\textwidth]{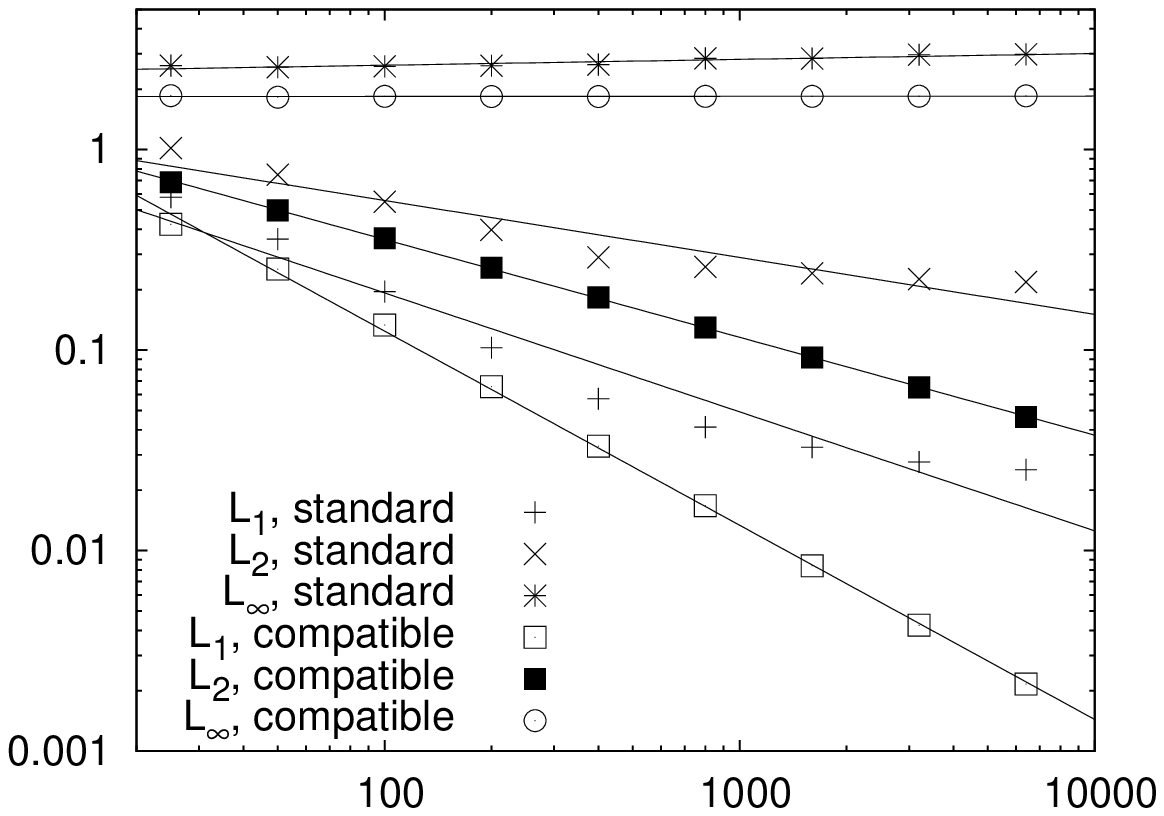}
} & 
\subfigure{
\includegraphics*[width=0.38\textwidth]{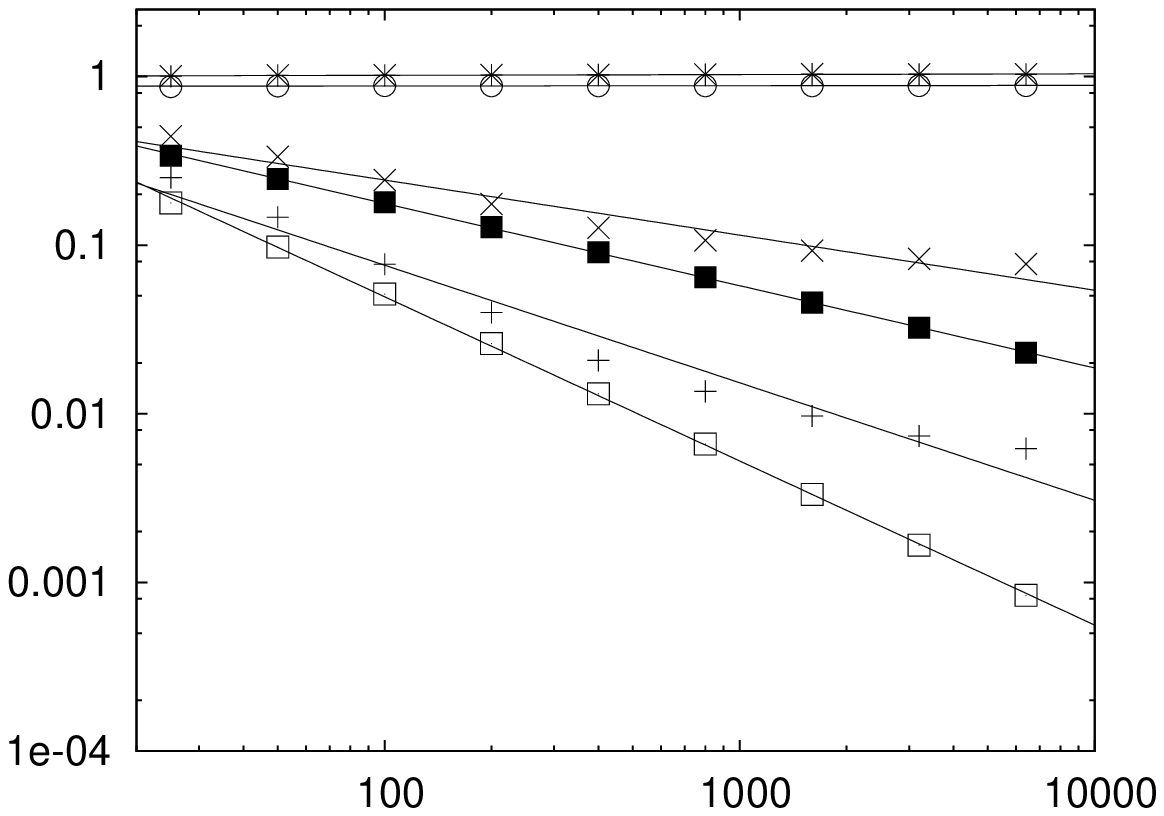}
} & 
\subfigure{
\includegraphics*[width=0.38\textwidth]{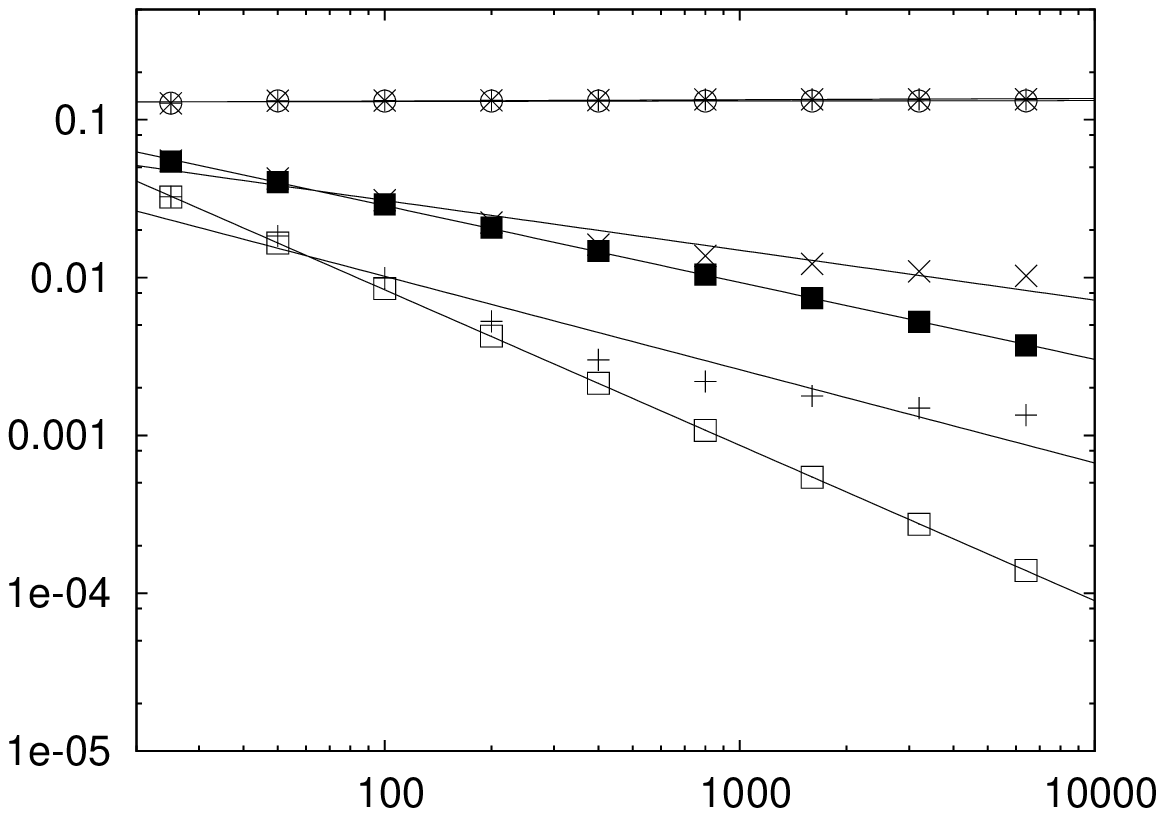}
}
\end{tabular}
\centerline{$N$}
\end{minipage}
\caption{Error estimates (points) and the fitted convergence rates (lines) for
  the mass density, velocity, and $A(s)$ in the 1-D planar Noh test problem.
}
\label{PlanarNohErrors.fig}
\end{figure}

\begin{table}
\begin{tabular}[htb]{|cr|c|c|c|} \hline
 &  & $L_1$ & $L_2$ & $L_\infty$ \\ \hline
 &  Std. & -0.59 $\pm$ 0.06 & -0.28 $\pm$ 0.03 & 0.0292 $\pm$ 0.004 \\
\raisebox{2.5ex}{$\rho$} 
 & Comp. & -0.97 $\pm$ 0.01 & -0.488 $\pm$ 0.002 & 0.001 $\pm$ 0.0001 \\ \hline
 &  Std. & -0.69 $\pm$ 0.05 & -0.33 $\pm$ 0.03 & 0.004 $\pm$ 0.0008 \\
\raisebox{2.5ex}{$v$} 
 & Comp. & -0.974 $\pm$ 0.007 & -0.488 $\pm$ 0.002   & 0.0016 $\pm$ 0.0006 \\ \hline
 &  Std. & -0.59 $\pm$ 0.06 & -0.32 $\pm$ 0.03 & 0.008 $\pm$ 0.002 \\
\raisebox{2.5ex}{$A(s)$} 
 & Comp. & -0.985 $\pm$ 0.002 & -0.487 $\pm$ 0.003  & 0.003 $\pm$ 0.002 \\ \hline
\end{tabular}
\caption{Fitted convergence rates for the mass density, velocity, and $A(s)$ in the
  planar Noh problem, shown $\pm 1 \sigma$.
}
\label{PlanarNohConv.tab}
\end{table}
We now consider a strong shock test case, the planar Noh problem \cite{Noh87}.
In this test case two streams of initially pressureless gas collide at the
origin, setting up an infinite strength shock propagating back into the gas
streams, behind which the gas is hot and motionless.  We again consider a
$\gamma = 5/3$ gas, with initial conditions $(\rho, P, v) = (1, 0, -1)$ for $x >
0$ and a reflecting boundary condition at $x = 0$.  We run simulations with $N
\in$ (25, 50, 100, 200, 400, 800, 1600, 3200, \& 6400) nodes.

Figure \ref{PlanarNohProfiles.fig} shows the profiles of the mass density,
velocities, and $A(s)$ at $t = 0.3$.  The shock speed for our initial conditions
is $v_s = 1/3$, so the expected shock position is 0.1.  The erroneous dip in the
mass densities (and spike in $A(s)$) near the origin is the so called ``wall
heating'' effect.  It appears the compatible discretization has slightly less
wall heating than the standard formalism (the dip in the mass density is less),
as well as a smaller overshoot in $A(s)$ in this region.

Examination of the error norms in Figure \ref{PlanarNohErrors.fig} and the
fitted convergence rates summarized in Table \ref{PlanarNohConv.tab} shows that
this strong shock test problem definitely benefits from the compatible
discretization.  The standard formalism fails to achieve first order convergence
(hovering in the range $m \in [-0.6, -0.7]$), while the compatible formalism
does see first order.  This results in the compatible discretization achieving
better accuracy for the same number of points as we increase the resolution.
The critical difference between this problem and previous shock tube test case
is most likely that here the energy is converted entirely from kinetic to
thermal energy through the shock, while the weaker shock in the Sod test does
not convert nearly as much of the energy budget in the problem between these
forms.  The exact energy conservation guaranteed by the compatible
discretization ensures the work incurred by the accelerations due to the
artificial viscosity (the dominant numerical term for this problem) is
accurately represented, leading to the improved accuracy noted here.  Close
examination of the the profiles from all the simulations reveals that the major
source of the error in the standard case is the shock position, which is
slightly ahead of where it should be.  This is mostly likely related to the fact
that the standard form suffers a slight growth in the energy, with $\Delta E/E
\sim 0.003$ at these resolutions.  Although conservation to less than half a
percent is typically considered quite reasonable, in this case we can measure
the effects of that error in the final solution.

\subsection{Cylindrical Noh shock test}
\label{NohCyl.sec}
In this section we look at the only multi-dimensional test considered in this
paper: the cylindrical Noh problem \cite{Noh87} modeled in 2-D.  This is largely
presented as a demonstration that there is nothing special about the 1-D tests
we have considered so far, and the compatibly differenced formalism works as
well in multi-dimensional problems as in 1-D.  The cylindrical Noh test posits
that we have a cylindrically convergent flow of an initially pressureless,
uniform density gas.  In cylindrical coordinates $(r, \theta, z)$ the initial
conditions we use can be expressed as $(\rho, P, v_r, v_\theta, v_z) = (1, 0,
-1, 0, 0)$.  We again use a $\gamma = 5/3$ gamma-law gas, so the post-shock
density should be a constant factor of 16, with the gas adiabatically compressed
in the pre-shock region by a factor of 4 and then shocked an additional factor
of 4 due to the shock transition.  We initialize the problem with the SPH nodes
on rings (with azimuthal spacing set equal to the initial radial spacing), and
consider problems with the number of radial rings $N_r \in$ (25, 50, 100, 200,
\& 400).

It is worth noting that in general multi-dimensional tests such as this require
the anisotropic sampling of ASPH \cite{asph98} in order to achieve the expected
convergence rates.  However, this case is special because in 2-D the appropriate
post-shock compression is in fact isotropic, so the circular shape of the 2-D
SPH smoothing scale is the appropriate solution.  In the pre-shock region there
is anisotropic compression in the azimuthal direction, but this does not
compromise the radial resolution significantly and therefore does not adversely
affect the results.  We do however require the tensor artificial viscosity
\cite{TensorQ} with it's associated limiter to get reasonable results, which is
why we include this term in Eqs.\ (\ref{tsphmom.eq}) \& (\ref{tsphenergy.eq}).

\begin{figure}[htb]
\begin{minipage}[b]{\textwidth}
\begin{tabular}[h]{rccc}
 & $\rho$ & $v_r$ & $A(s)$ \\
\turnbox{90}{\hspace{0.05\textwidth}Standard} &
\subfigure{
\includegraphics*[width=0.35\textwidth]{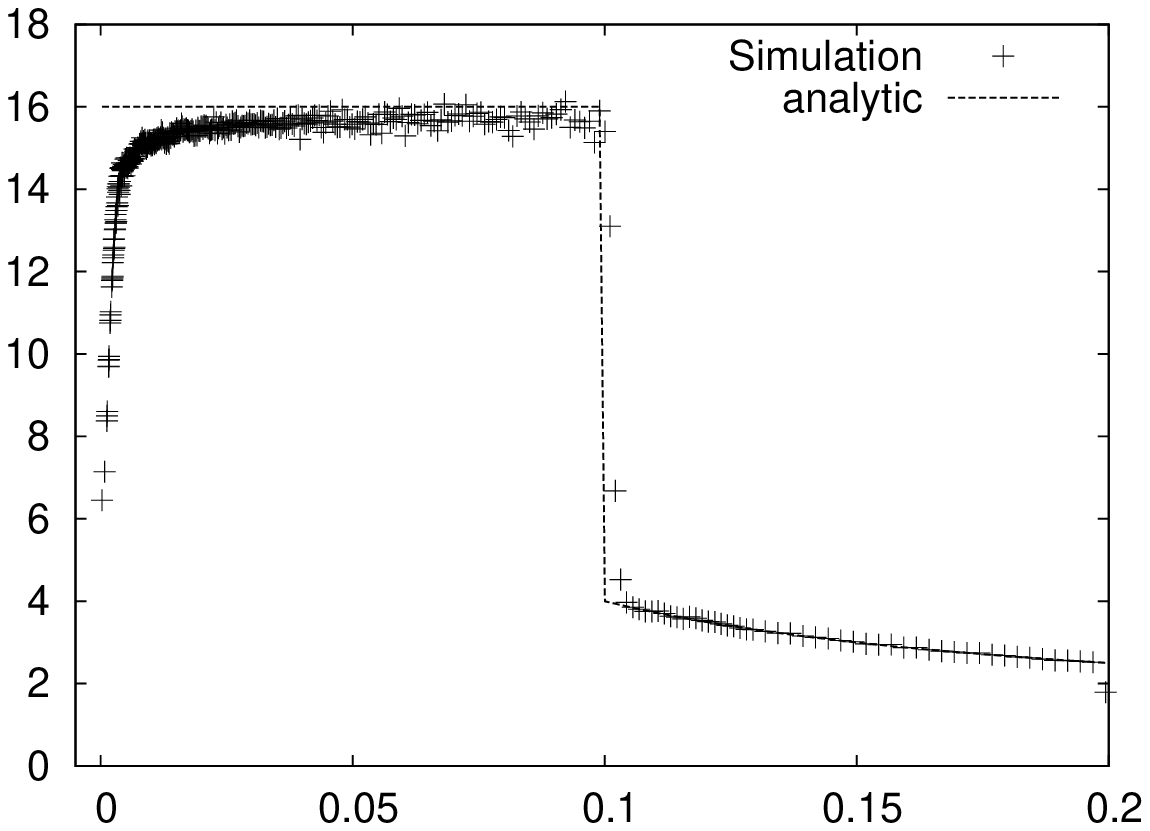}
} &
\subfigure{
\includegraphics*[width=0.35\textwidth]{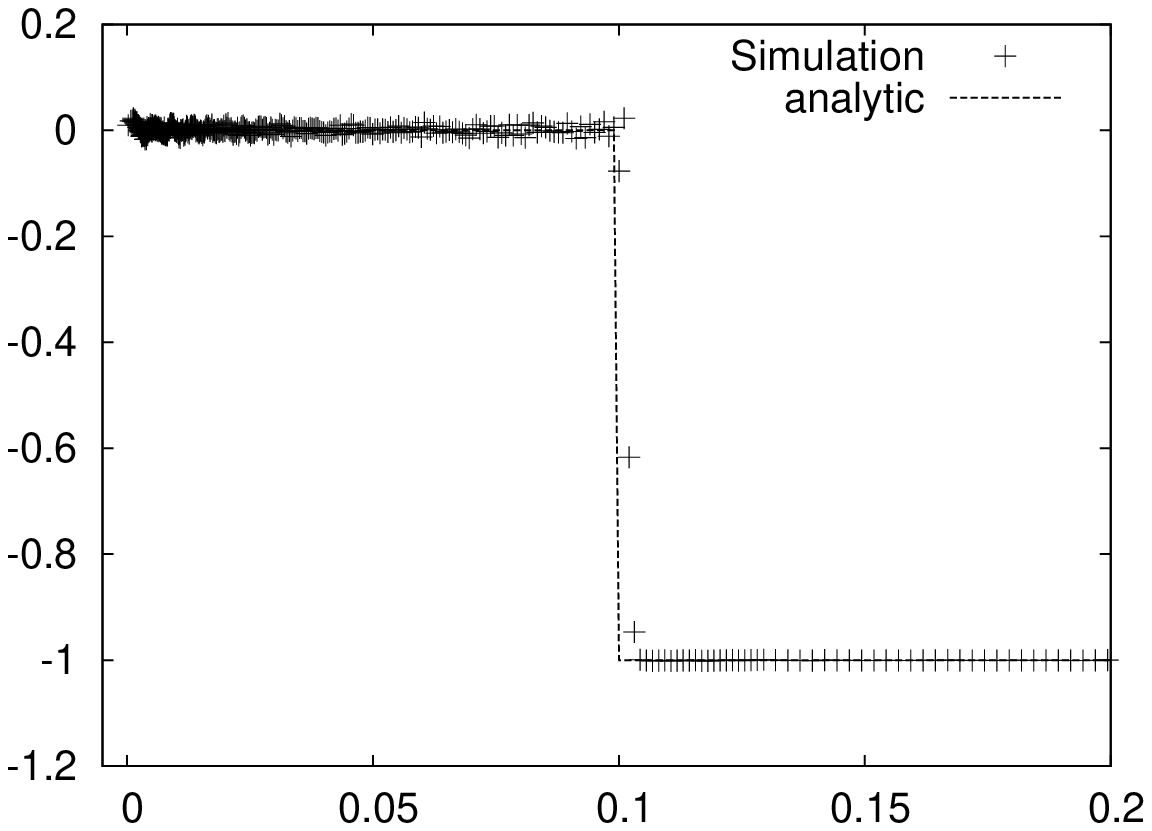}
} &
\subfigure{
\includegraphics*[width=0.35\textwidth]{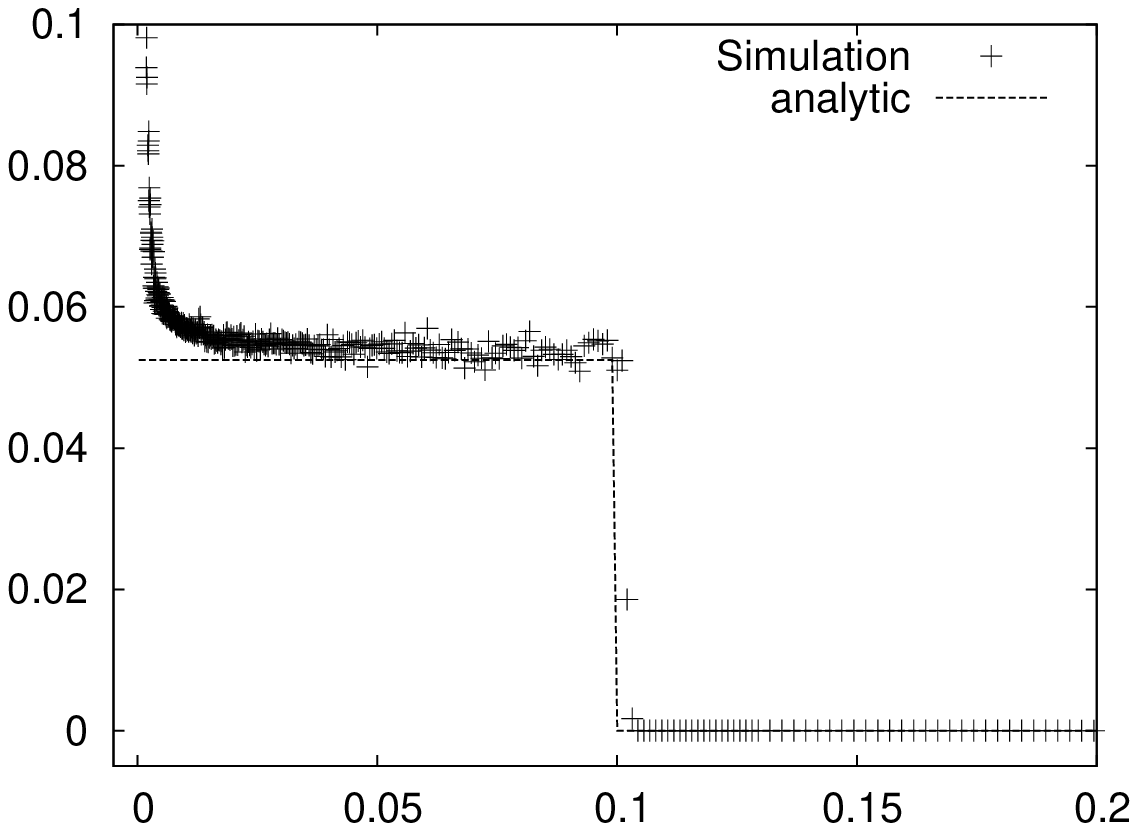}
} \\
\turnbox{90}{\hspace{0.1\textwidth}Compatible} &
\subfigure{
\includegraphics*[width=0.35\textwidth]{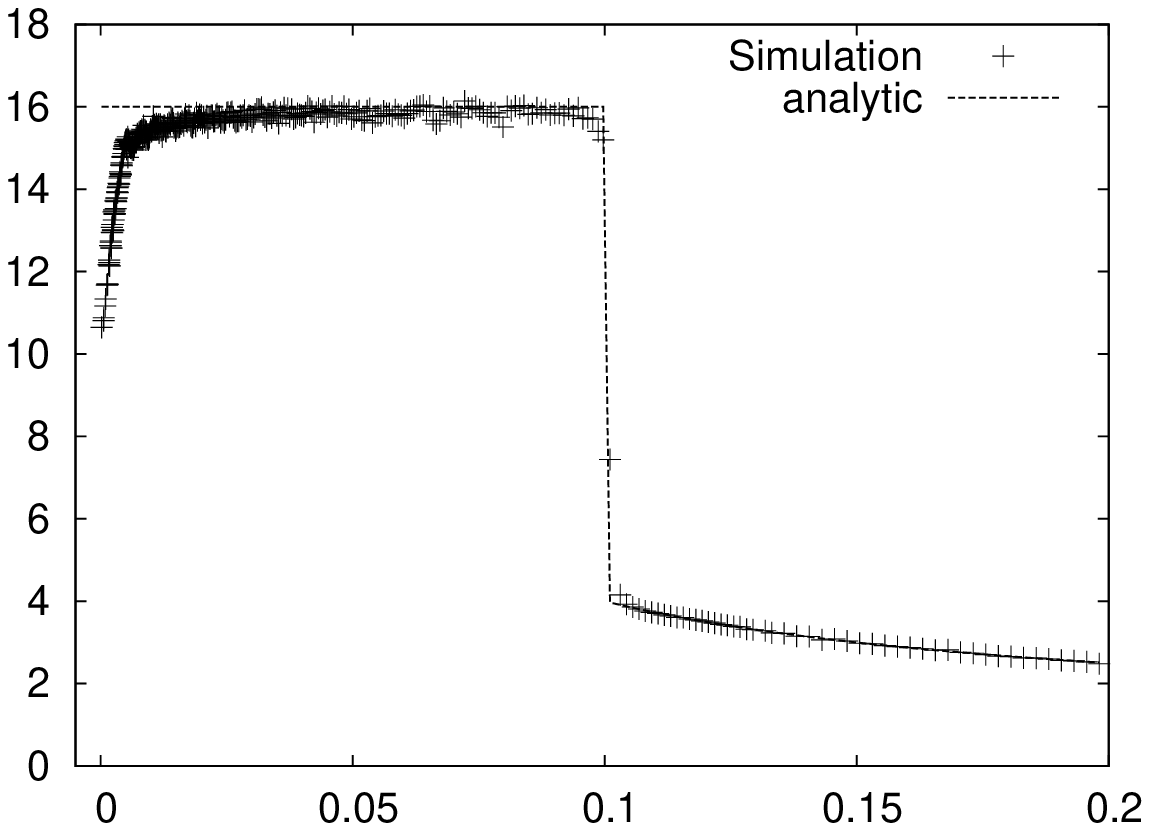}
} &
\subfigure{
\includegraphics*[width=0.35\textwidth]{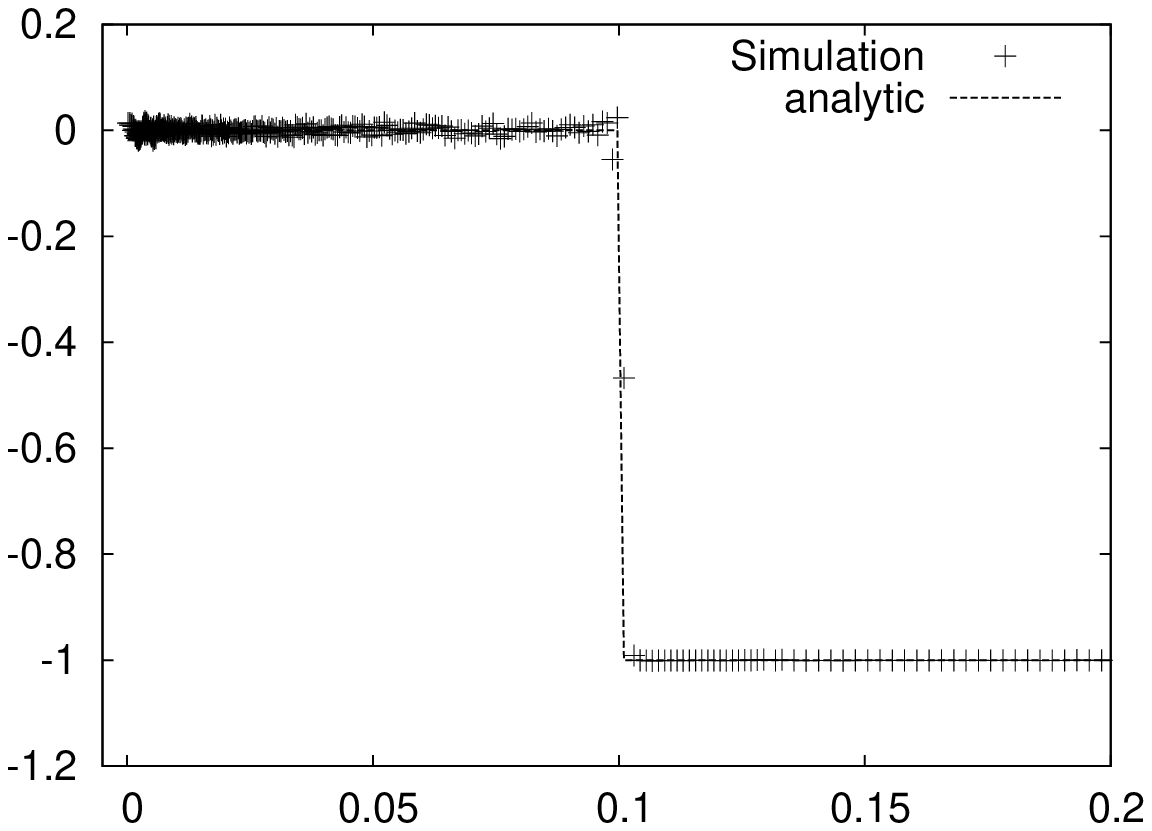}
} &
\subfigure{
\includegraphics*[width=0.35\textwidth]{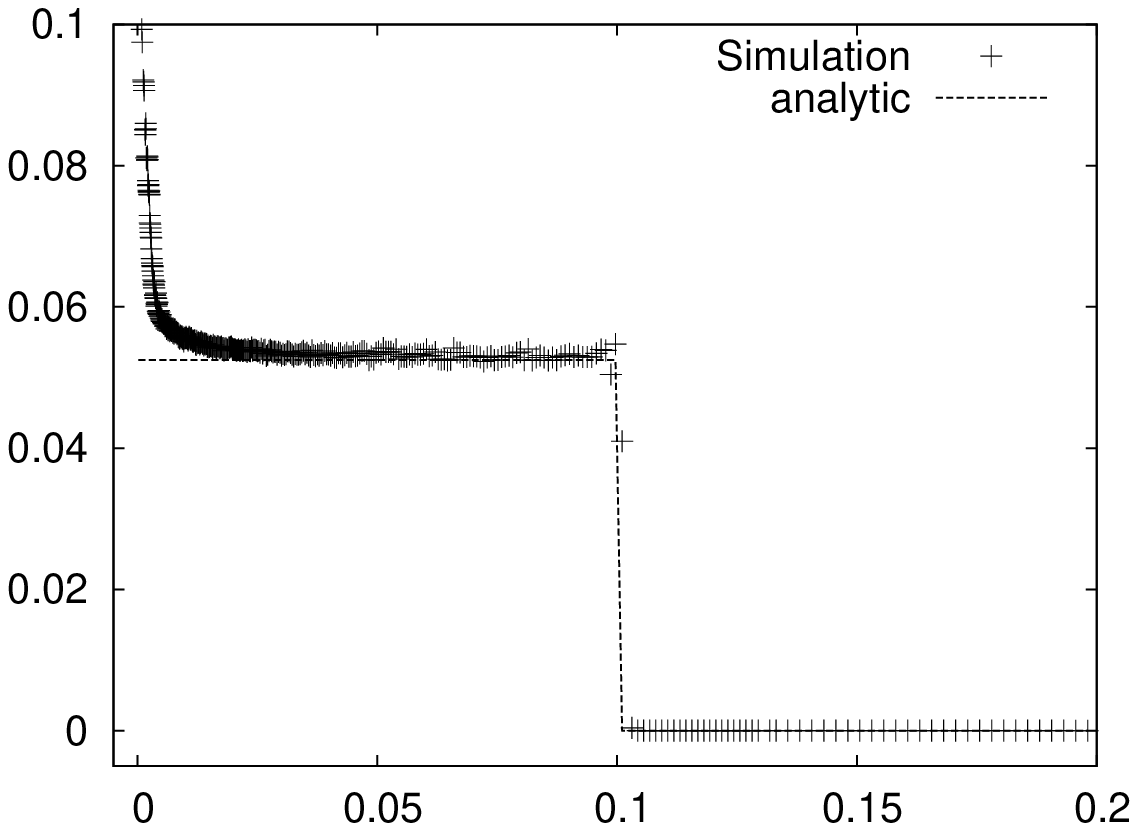}
}
\end{tabular}
\centerline{$x$}
\end{minipage}
\caption{Radial profiles of the mass density, radial velocity, and $A(s)$ for the
  cylindrical Noh problem.  Shown here are the results for the simulation
  using 400 radial rings of nodes (points) vs.\ the analytic solution (lines).
}
\label{CylNohProfiles.fig}
\end{figure}
Examination of the profiles of the mass density, velocity, and $A(s)$ in Figure
\ref{CylNohProfiles.fig} shows that the compatible formalism sees slightly less
wall heating and less of a corresponding spike in $A(s)$ compared with the
standard results, much like in the planar Noh problem.  Both methods match the
pre-shock conditions quite well.  However, it is evident that the compatible
discretization shows improvements in the post-shock profiles, in that it more
closely matches the post-shock density of 16, and $A(s)$ is closer to the
analytic answer and shows less scatter.  The improved post-shock density for the
compatible form appears to be related to the fact that the shock position is
again a little too far advanced in the run using the standard formalism.  In terms
of energy conservation we find that the standard form fares about as well as in
the planar case, with an energy drift $\Delta E/E \sim 0.004$ (the compatible
discretization maintains the expected $\Delta E/E \sim 10^{-14}$).

\begin{figure}[htb]
\begin{minipage}[b]{\textwidth}
\begin{tabular}[h]{rccc}
 & $\rho$ & $v_r$ & $A(s)$ \\
\turnbox{90}{\hspace{0.1\textwidth}Error} &
\subfigure{
\includegraphics*[width=0.38\textwidth]{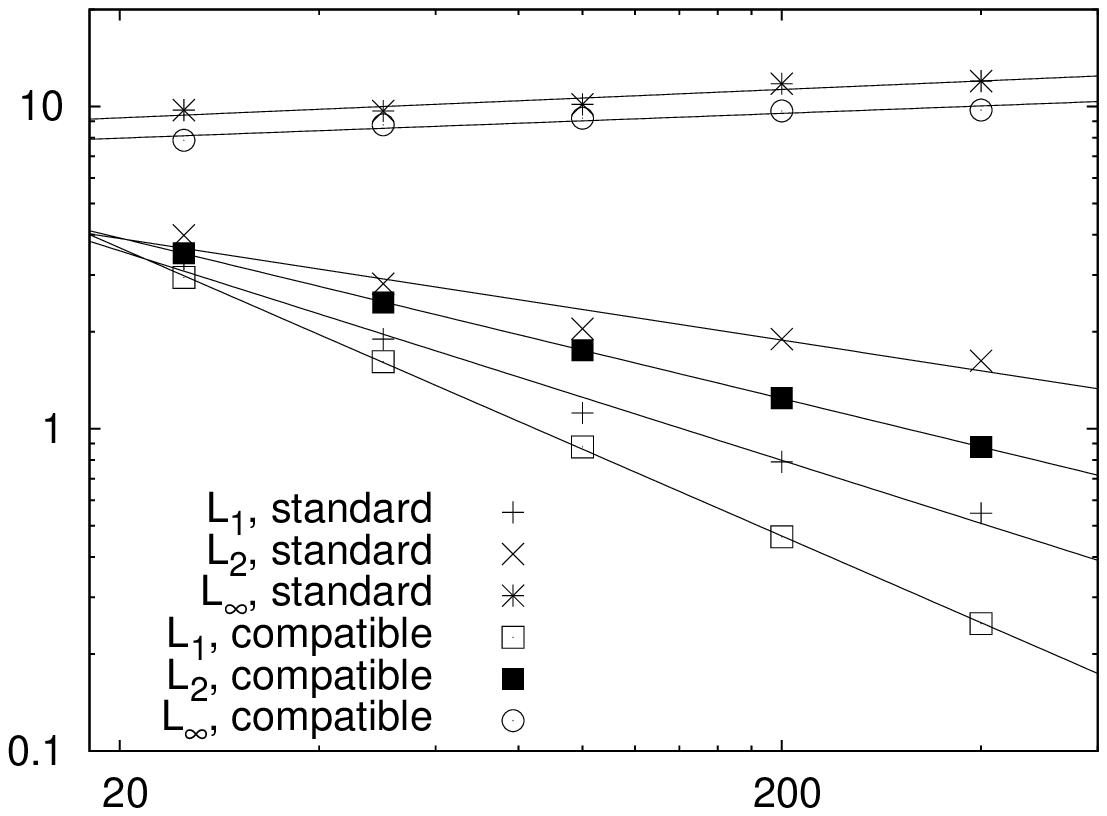}
} & 
\subfigure{
\includegraphics*[width=0.38\textwidth]{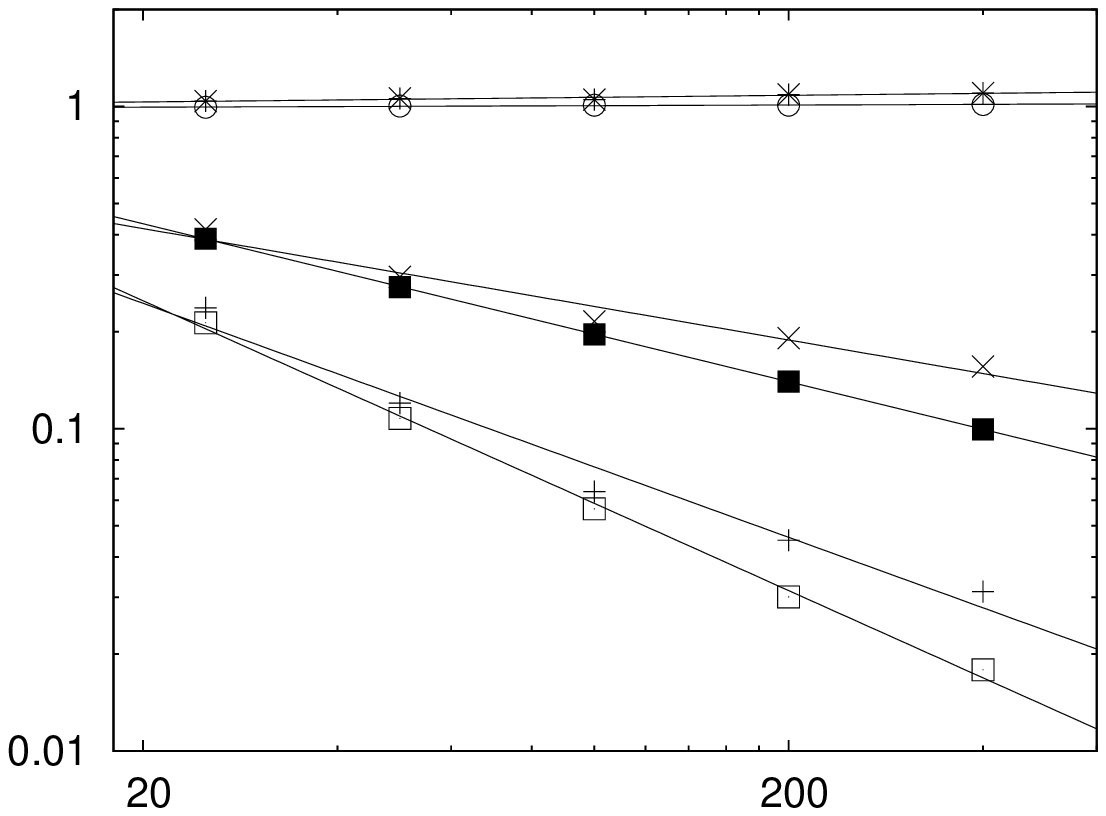}
} & 
\subfigure{
\includegraphics*[width=0.38\textwidth]{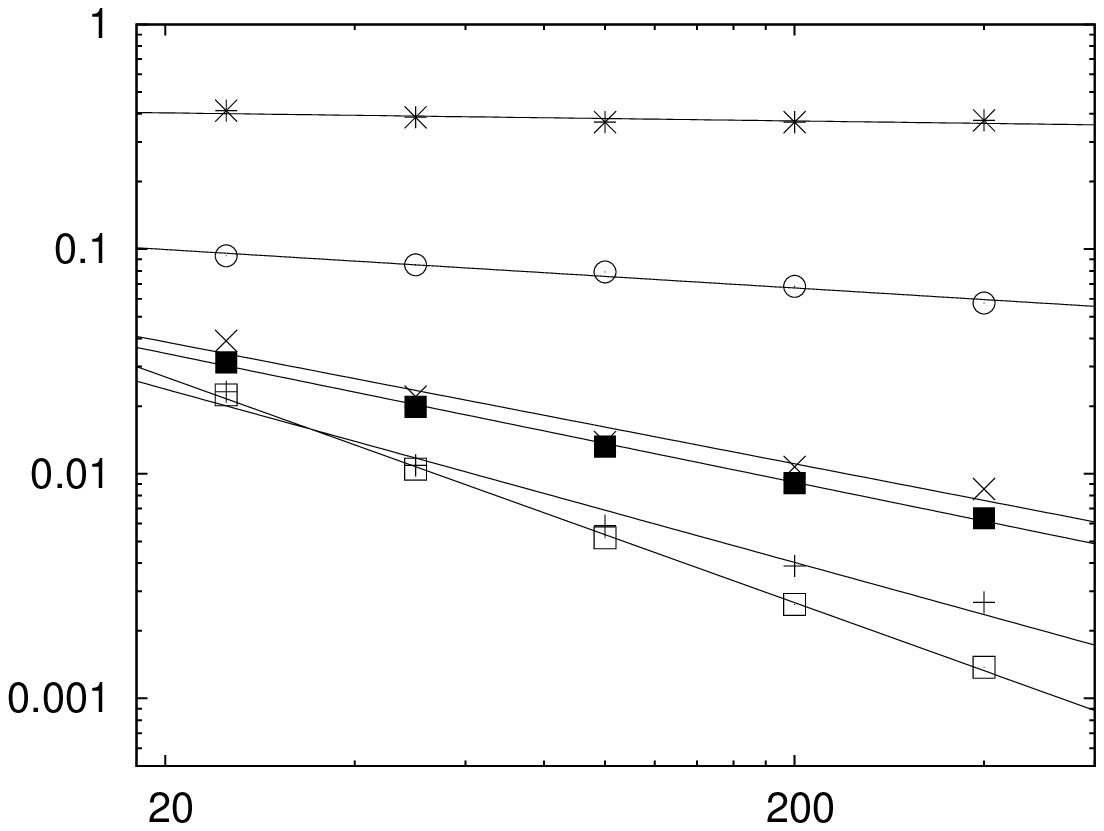}
}
\end{tabular}
\centerline{$N_r$}
\end{minipage}
\caption{Error estimates (points) and the fitted convergence rates (lines) for
  the mass density, radial velocity, and $A(s)$ in the 2-D cylindrical Noh
  test problem.
}
\label{CylNohErrors.fig}
\end{figure}

\begin{table}
\begin{tabular}[htb]{|cr|c|c|c|} \hline
 &  & $L_1$ & $L_2$ & $L_\infty$ \\ \hline
 &  Std. & -0.65 $\pm$ 0.04 & -0.32 $\pm$ 0.05 & 0.09 $\pm$ 0.02 \\
\raisebox{2.5ex}{$\rho$} 
 & Comp. & -0.894 $\pm$ 0.006 & -0.498 $\pm$ 0.002 & 0.08 $\pm$ 0.01 \\ \hline
 &  Std. & -0.73 $\pm$ 0.07 & -0.34 $\pm$ 0.04 & 0.020 $\pm$ 0.005 \\
\raisebox{2.5ex}{$v_r$} 
 & Comp. & -0.90 $\pm$ 0.02 & -0.490 $\pm$ 0.001   & 0.007 $\pm$ 0.001 \\ \hline
 &  Std. & -0.77 $\pm$ 0.07 & -0.54 $\pm$ 0.06 & -0.04 $\pm$ 0.02 \\
\raisebox{2.5ex}{$A(s)$} 
 & Comp. & -1.00 $\pm$ 0.02 & -0.57 $\pm$ 0.02  & -0.17 $\pm$ 0.02 \\ \hline
\end{tabular}
\caption{Fitted convergence rates for the mass density, radial velocity, and
  $A(s)$ in the cylindrical Noh problem, shown $\pm 1 \sigma$.
}
\label{CylNohConv.tab}
\end{table}
Examination of the measured errors and convergence rate in Figure
\ref{CylNohProfiles.fig} and Table \ref{CylNohConv.tab} again mirrors what we
saw in the planar Noh problem.  The standard discretization does not quite make
the expected first order convergence (hovering in the range $m \in [-0.65, -0.8]$),
while the compatible discretization does maintain first order convergence.

\subsection{Planar Taylor-Sedov blastwave test}
\label{SedovPlanar.sec}
The final test case we consider is the Taylor-Sedov blastwave \cite{Sedov59},
which is the problem of an intense explosion due to the injection of a point
source of energy into an otherwise uniform, pressureless gas.  This test is a
useful diagnostic of how an algorithm handles relaxing extreme discontinuities
in the specific thermal energy -- in particular, as discussed earlier particular
choices for the partitioning of the work between nodes can result in
unphysically negative specific thermal energies.  We model this problem in 1-D,
corresponding to the planar solution in \cite{Sedov59}.  We use a $\gamma = 5/3$
gamma-law gas, and consider runs using $N \in$ (51, 101, 201, 401, 801, 1601,
3201, 6401, 12801, \& 25601) SPH nodes seeded in the volume $x \in [-1,1]$.  Our
initial conditions are $(\rho, P, v) = (1, 0, 0)$, and we seed an energy spike
corresponding to $E_{\mbox{\scriptsize spike}} = 1$ on the single point at the
origin.

\begin{figure}[htb]
\begin{minipage}[b]{\textwidth}
\begin{tabular}[h]{rccc}
 & $\rho$ & $v$ & $A(s)$ \\
\turnbox{90}{\hspace{0.05\textwidth}Standard} &
\subfigure{
\includegraphics*[width=0.35\textwidth]{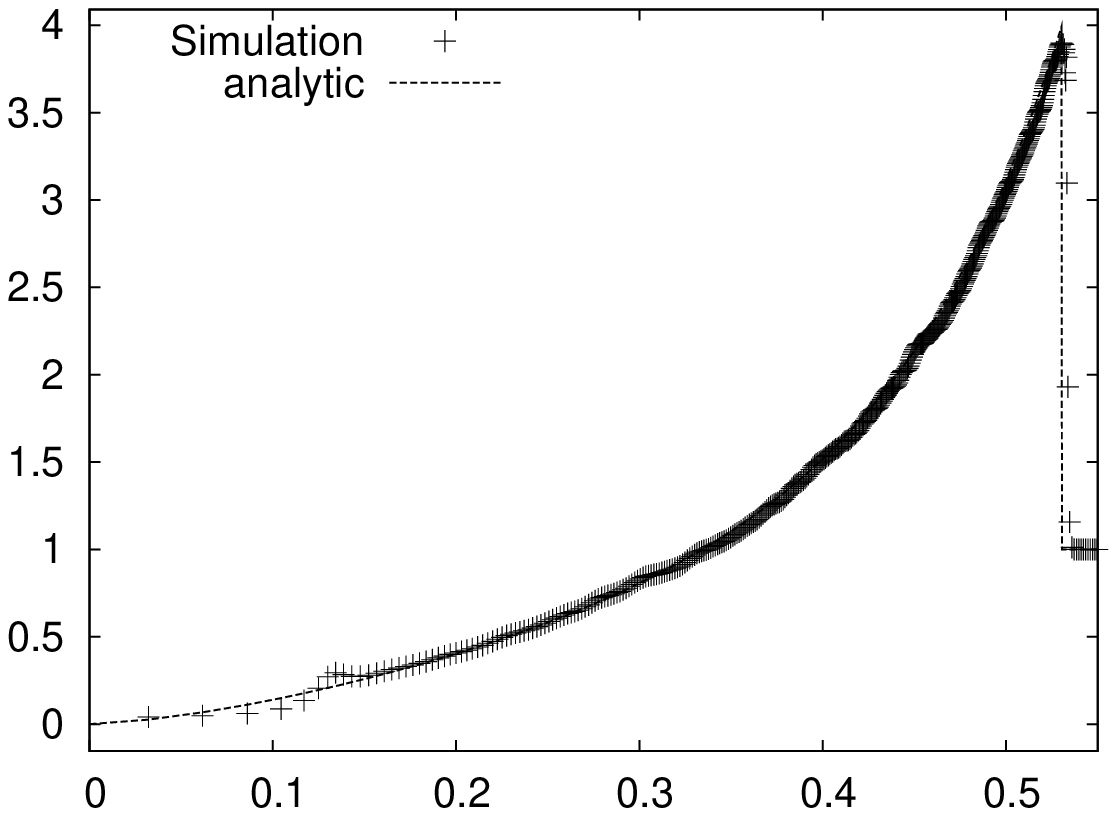}
} &
\subfigure{
\includegraphics*[width=0.35\textwidth]{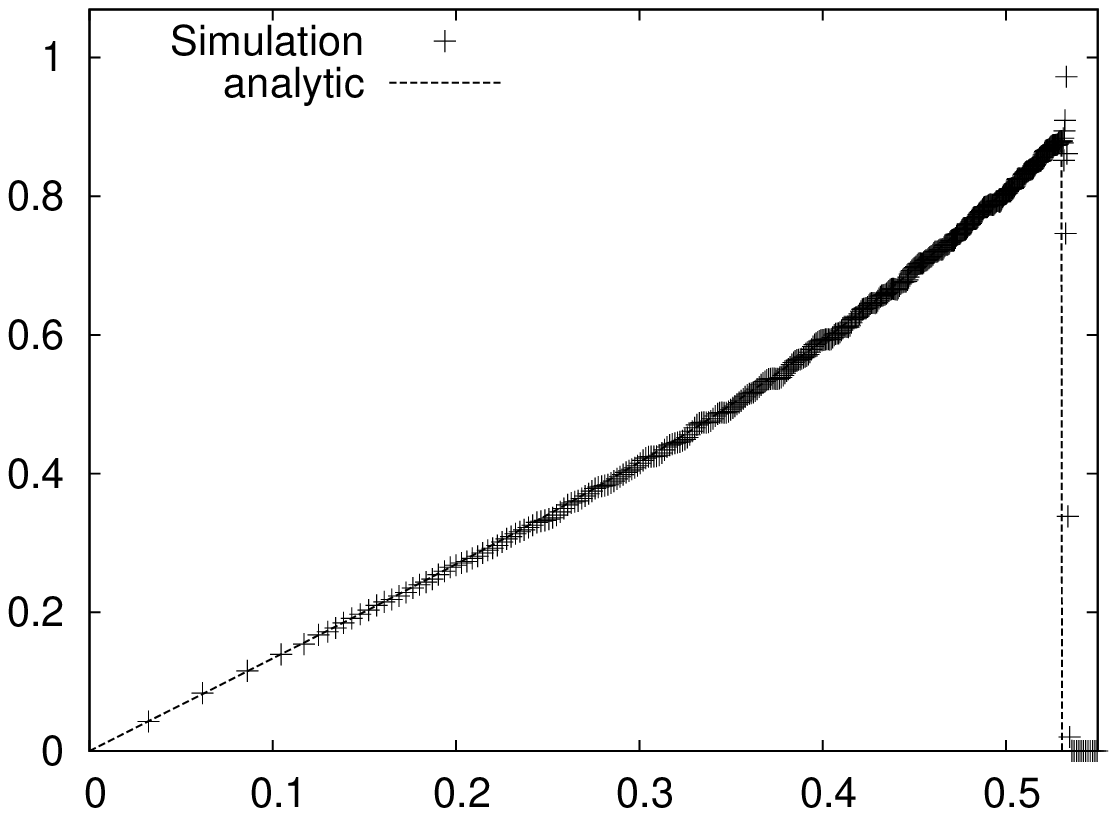}
} &
\subfigure{
\includegraphics*[width=0.35\textwidth]{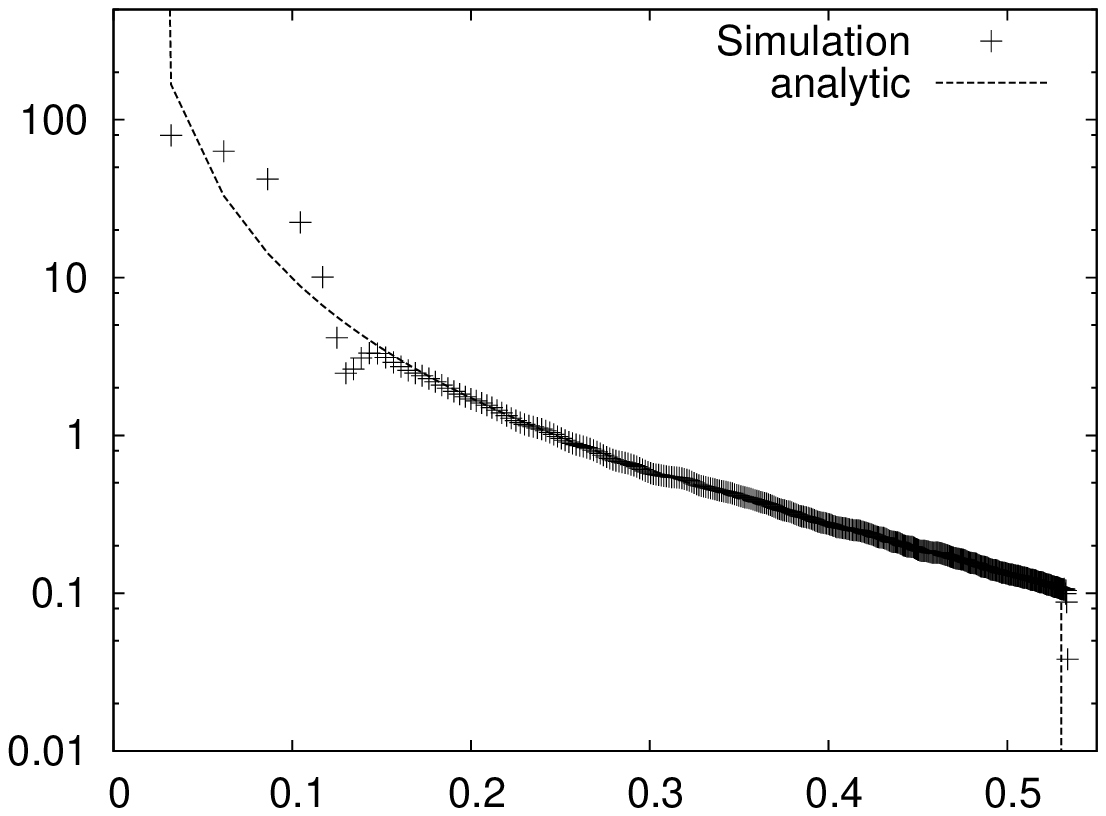}
} \\
\turnbox{90}{\hspace{0.1\textwidth}Compatible} &
\subfigure{
\includegraphics*[width=0.35\textwidth]{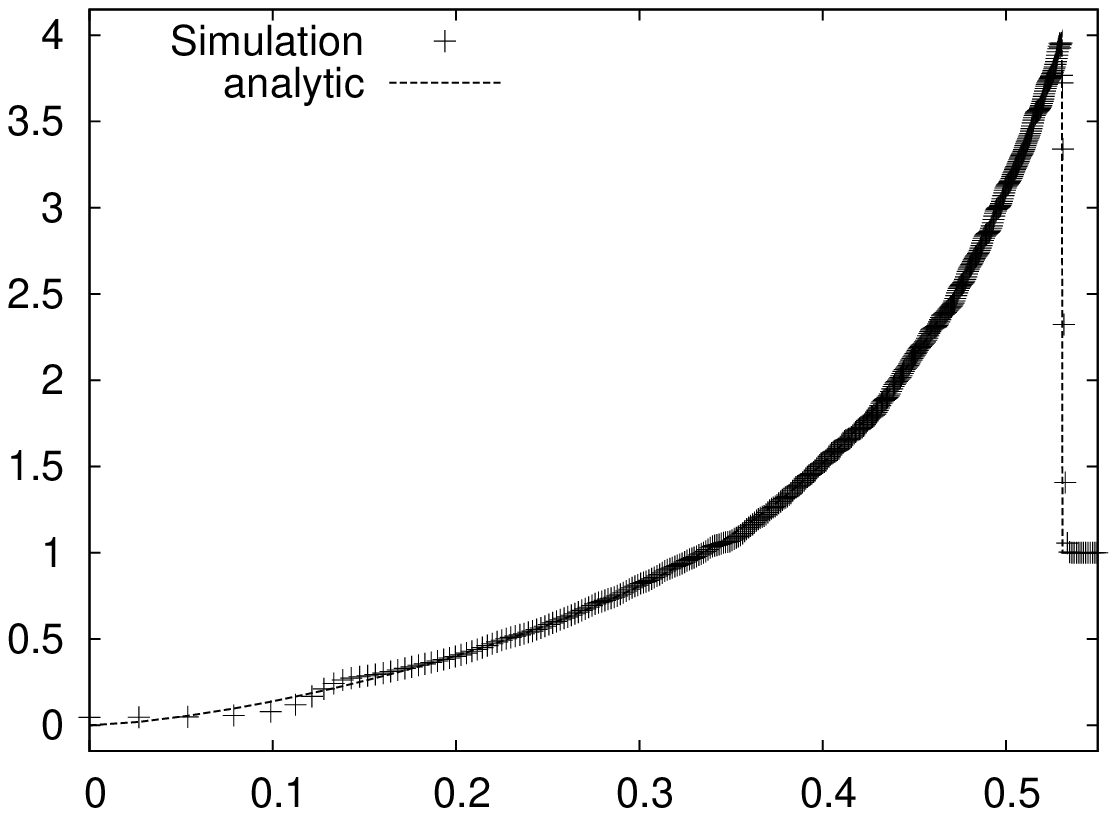}
} &
\subfigure{
\includegraphics*[width=0.35\textwidth]{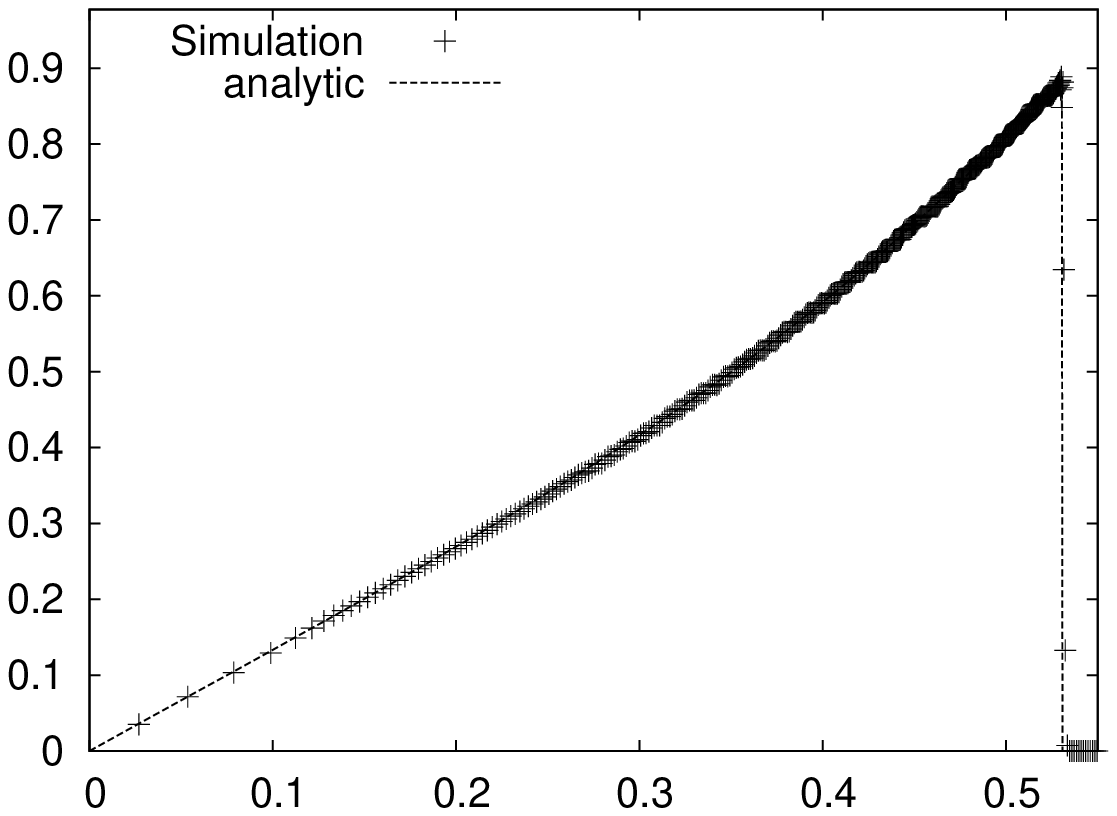}
} &
\subfigure{
\includegraphics*[width=0.35\textwidth]{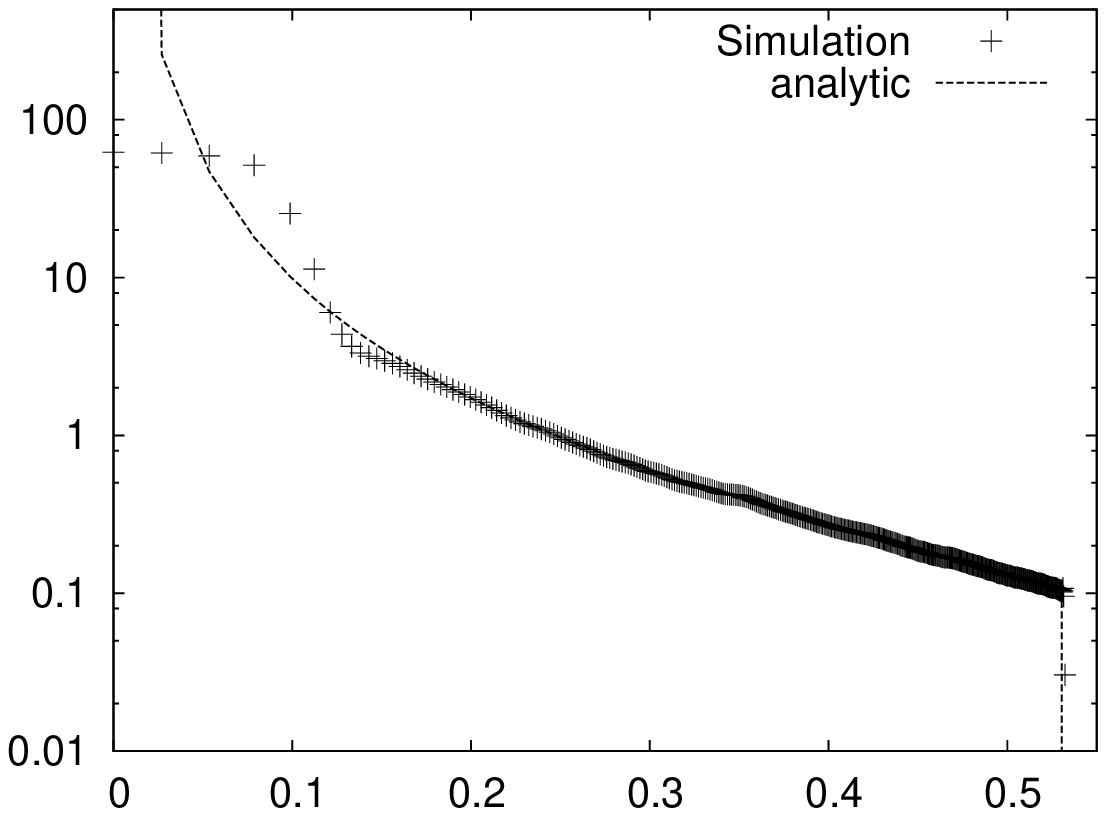}
}
\end{tabular}
\centerline{$x$}
\end{minipage}
\caption{Profiles of the mass density, velocity, and $A(s)$ for the planar
  Taylor-Sedov blastwave test case.  Shown here are results from the $N = 801$
  node simulations (points) compared with the analytic solution (lines).
}
\label{SedovProfiles.fig}
\end{figure}
\begin{figure}[htb]
\begin{minipage}[b]{\textwidth}
\begin{tabular}[h]{rcc}
 & Standard & Compatible \\
\turnbox{90}{\hspace{0.2\textwidth}$\rho$} &
\subfigure{
\includegraphics*[width=0.5\textwidth]{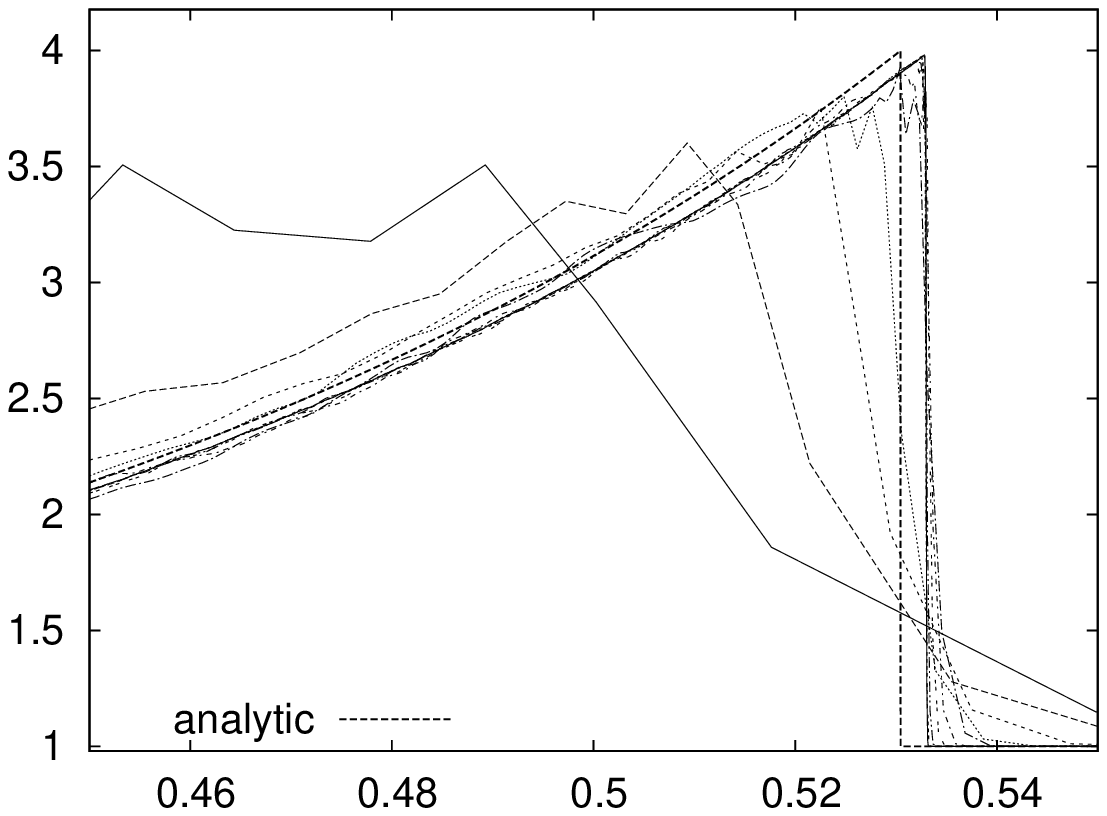}
}
&
\subfigure{
\includegraphics*[width=0.5\textwidth]{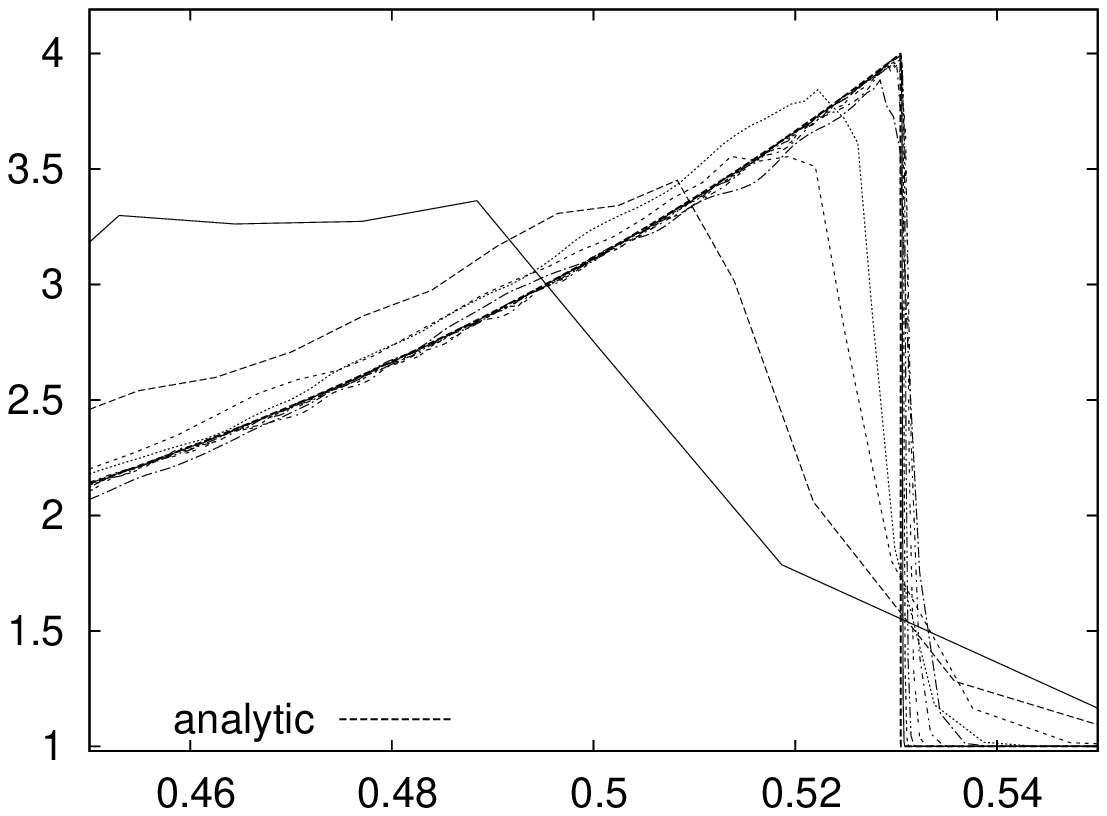}
}
\end{tabular}
\centerline{$x$}
\end{minipage}
\caption{Zoom in on the details of the mass density profiles around the shock
  position in the Taylor-Sedov blastwave test case.  Here we have plotted the
  profiles from all the resolutions ($N$ ranging from 51 to 25601) to see how
  the profiles are converging as compared with the analytic solution.
}
\label{SedovZoomProfiles.fig}
\end{figure}
Figure \ref{SedovProfiles.fig} shows the profiles of the mass density, velocity,
and $A(s)$ at $t$ = 0.3 for $x > 0$, while Figure \ref{SedovZoomProfiles.fig}
zooms in on the mass density profiles around the shock position at this same
time.  It is immediately clear that in this problem the standard discretization
is converging on the wrong shock position, although it is not far off.  The
false convergence point is just outside the expected value from the analytic
solution, and the fact that the converged shock position is a bit too far out is
consistent with the fact that the energy grows slightly, with $\Delta E/E \sim$
0.1\%--1\%, increasing slowly with resolution.  Interestingly, in running
this problem we found that it was necessary to smooth the initial energy spike
using a Shephard's function formalism
\beq
   E_i = \frac{E_{\mbox{\scriptsize spike}} W(x_i - x_0, h_i)}
              {\sum_j W(x_j - x_0, h_j)},
\eeq
(where $E_{\mbox{\scriptsize spike}}, x_0$ are the energy and position of the
energy spike) in order to get either formalism to converge to the correct
position.  Without this smoothing, the simulated shock was too slow, and the
shock position consistently lagged behind the analytic expectation.  Note that
this smoothing has nothing to do with avoiding negative internal energies -- both
formalisms used here successfully run this problem without generating any
negative energies.  Rather, we found this initial smoothing step necessary
solely to improve the match to the position of the shock.  Most likely the
reason this is not typically seen in SPH simulations of this problem is due to
the fact that usually those models are not run at sufficient resolution to
distinguish the converged shock position as is done here.  In terms of the
post-shock profiles, to the eye both schemes are doing a reasonable job of
capturing the post-shock shapes of the mass density, velocity, and $A(s)$.  They
each have some difficulty capturing the near vacuum conditions at the origin
(particularly notable in $A(s)$) because this is a Lagrangian scheme, and we can
only represent a solution where there is mass.

\begin{figure}[htb]
\begin{minipage}[b]{\textwidth}
\begin{tabular}[h]{rccc}
 & $\rho$ & $v$ & $A(s)$ \\
\turnbox{90}{\hspace{0.1\textwidth}Error} &
\subfigure{
\includegraphics*[width=0.38\textwidth]{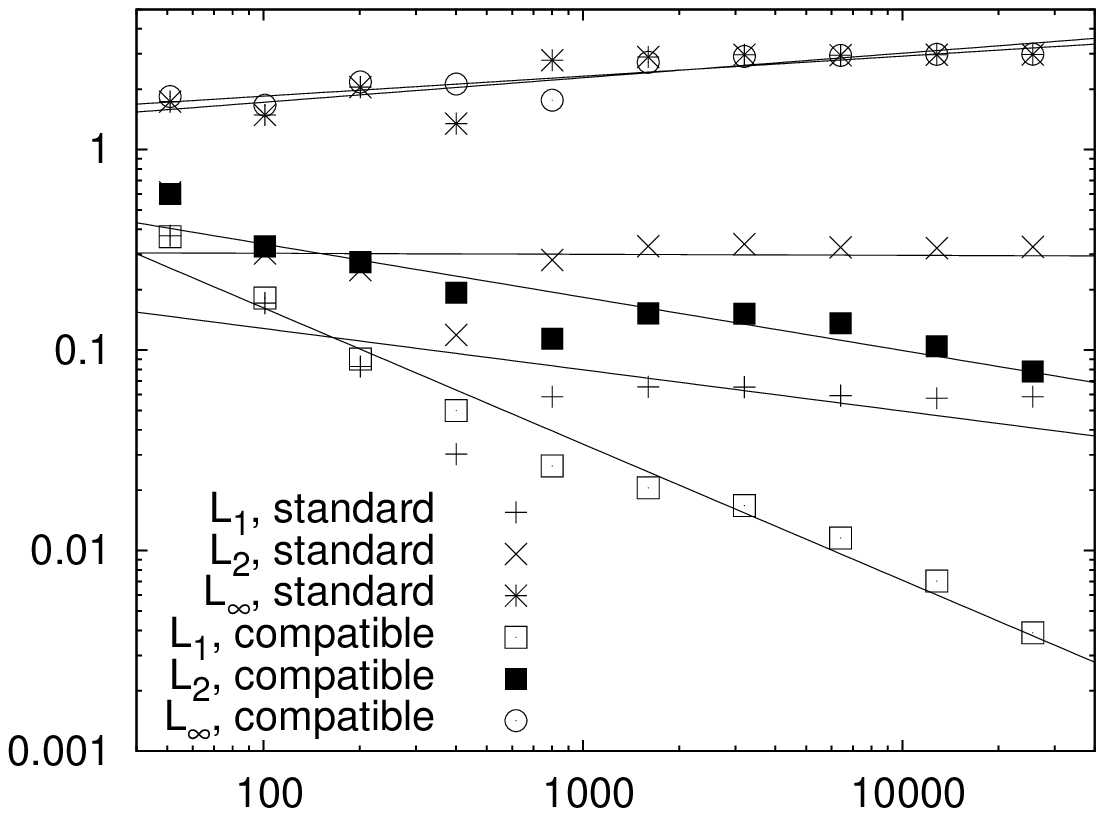}
} & 
\subfigure{
\includegraphics*[width=0.38\textwidth]{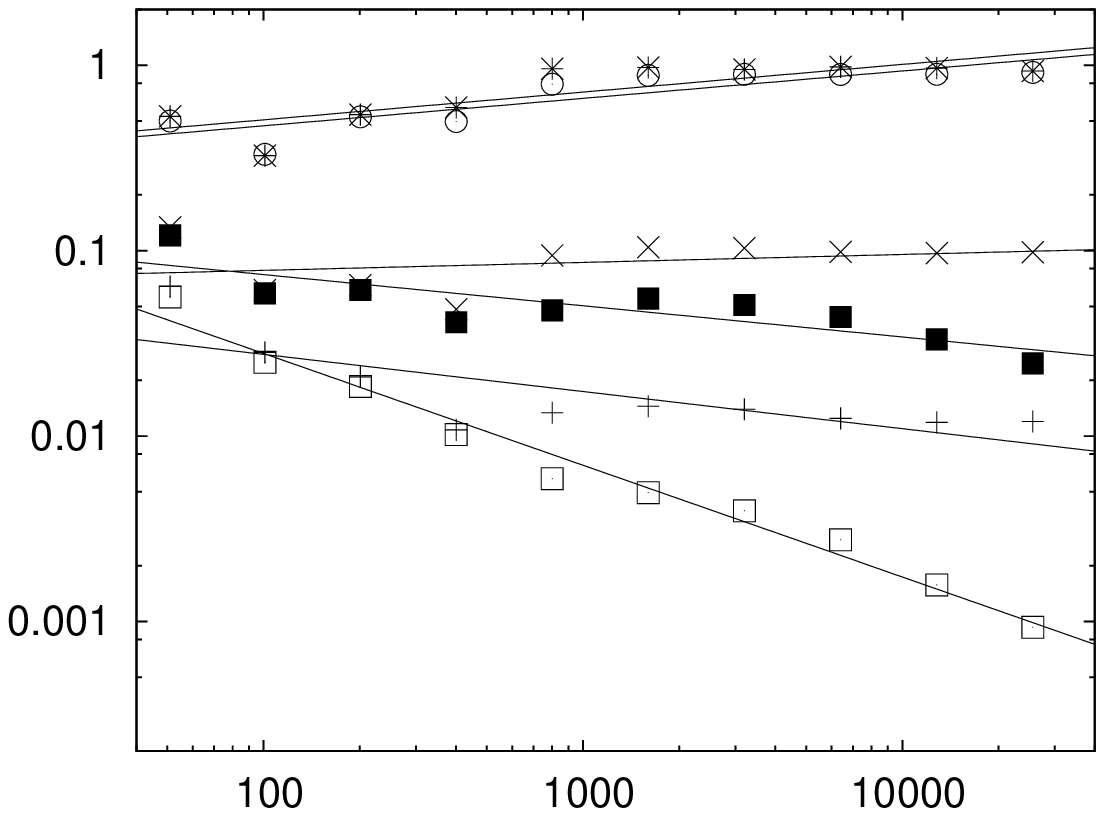}
} & 
\subfigure{
\includegraphics*[width=0.38\textwidth]{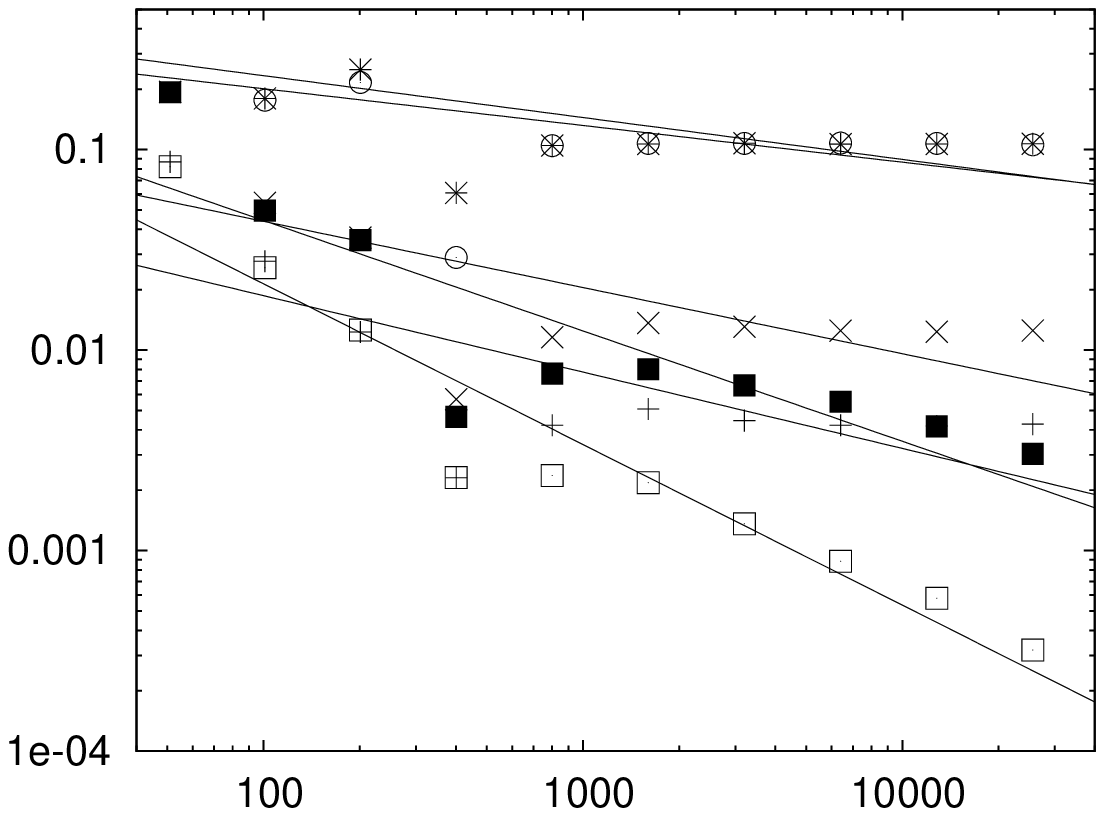}
}
\end{tabular}
\centerline{$N$}
\end{minipage}
\caption{Error estimates (points) and the fitted convergence rates (lines) for
  the mass density, velocity, and $A(s)$ in the Taylor-Sedov blastwave test problem.
}
\label{SedovErrors.fig}
\end{figure}
\begin{table}
\begin{tabular}[htb]{|cr|c|c|c|} \hline
 &  & $L_1$ & $L_2$ & $L_\infty$ \\ \hline
 &  Std. & -0.24 $\pm$ 0.08 & -0.03 $\pm$ 0.06 & 0.12 $\pm$ 0.02 \\
\raisebox{2.5ex}{$\rho$} 
 & Comp. & -0.69 $\pm$ 0.04 & -0.28 $\pm$ 0.04 & 0.10 $\pm$ 0.02 \\ \hline
 &  Std. & -0.23 $\pm$ 0.06 & 0.04 $\pm$ 0.07 & 0.16 $\pm$ 0.05 \\
\raisebox{2.5ex}{$v$} 
 & Comp. & -0.62 $\pm$ 0.03 & -0.18 $\pm$ 0.04   & 0.15 $\pm$ 0.03 \\ \hline
 &  Std. & -0.4 $\pm$ 0.1 & -0.3 $\pm$ 0.1 & -0.2 $\pm$ 0.1 \\
\raisebox{2.5ex}{$A(s)$} 
 & Comp. & -0.80 $\pm$ 0.09 & -0.6 $\pm$ 0.1  & -0.2 $\pm$ 0.1 \\ \hline
\end{tabular}
\caption{Fitted convergence rates for the mass density, velocity, and $A(s)$ in the
  planar Taylor-Sedov blastwave, shown $\pm 1 \sigma$.
}
\label{SedovConv.tab}
\end{table}
The error norms and measured convergence rates in Figure \ref{SedovErrors.fig}
and Table \ref{SedovConv.tab} demonstrate the difficulty these schemes have in
matching the shock position.  The standard form sees quite poor convergence
rates (with $m \in [-0.2, -0.4]$), while the compatible formalism sees higher
convergence rates ($m \in [-0.6, -0.8]$) and corresponding more accurate
solutions, but still not quite the ideal first order.  If we play games such as
measuring the convergence by normalizing the positions of the profiles to the
measured shock position and compare to a normalized analytic answer, our
measured errors improve substantially and we see convergence at first order for
both schemes.  It is debatable whether it is more appropriate to fudge the error
measurements in this manner or not, but for this work we have elected to simply
measure the errors against the exact solution including the shock position,
since this is after all what we would be looking for were we modeling this
problem without access to the analytic solution.

\section{Conclusions}
\label{conclusions.sec}
We have derived a total energy conserving form of SPH based on the compatible
differencing ideas described in \cite{CompatibleHydro98}.  The compatible
formalism guarantees energy conservation by exactly accounting for the work done
by the discrete pair-wise accelerations between SPH nodes.  We describe a scheme
for dividing the pair-wise work between points such that the work term will tend
to drive the temperatures of the points closer to one and other, avoiding the
introduction of new extrema in the thermal energy (such as unphysical negative
temperature excursions).  This new scheme should represent a very minor
modification for existing SPH codes; indeed, as demonstrated in \S
\ref{fijstandard.sec} the algorithm described here is very closely related to
the standard discretizations of the energy equation.

We have demonstrated through a series of tests that the compatible energy
discretization for SPH results in quantifiable improvements in the accuracy and
convergence properties of the algorithm.  In order to address the natural
concern over the accuracy of the entropy solution in an energy conserving scheme
such as ours, we have explicitly investigated the accuracy of the entropy
evolution by examining the behaviour of the entropic function $A(s) =
P/\rho^\gamma$ in our tests, comparing the simulations to analytic solutions.
In all cases the accuracy of the entropy evolution has been improved by use of
the compatible discretization.  At first this might seem odd to someone
accustomed to seeing larger entropy errors in energy conserving schemes.  The
explanation comes from considering the differences in the approach taken here:
in this case we are still evolving the internal energy, we are simply making a
more accurate accounting of the discrete work done by the numerically estimated
accelerations.  Often energy conservation is achieved by evolving the total
energy and deriving the thermal energy by taking the difference between the
total and kinetic energies, which can squeeze the discretization errors into the
thermal term (particularly in strongly supersonic flows).  We view the
changes proposed here as more akin to the difference between updating the mass
density according the SPH sum (such as Eq.\ \ref{sphmass.eq}) vs.\ time evolving
the continuity equation.  In general using the sum definition for the density is
more accurate, because we have removed a source of error by not time evolving
the total mass in some sense.  There are times when evolving the continuity
equation makes more sense (such as solids with equations of state that are not
forgiving of the surface problems with Eq.\ \ref{sphmass.eq}), but in fluid
problems where the sum definition for the mass density is reasonable, it will
generally prove more accurate.  We believe this is analogous to the changes we
have proposed here.  By exactly accounting for the numerical work done by the
momentum equation (Eq.\ \ref{tsphmom.eq}), we have removed one of the components
of error (a time evolved definition of the energy) rather than just force the
discretization errors into the thermal energy equation.

We should also point out that we do not mean to suggest that the accuracy
improvements in the SPH discretization possible by accounting for the variable
$h$ terms such as described in \cite{EntropySPH02} and \cite{PriceMon06} should
be neglected.  These authors find that the fundamental accuracy of the numerics
is improved when self-consistently accounting for the effects of allowing a
variable smoothing scale, which also has the effect of improving the energy
conservation.  This improved energy conservation is simply a reflection of the
fact that the authors are achieving a more accurate solution of the problem for
some additional computational cost.  We would propose that one could see even
greater benefit by combining these approaches -- there is no reason one cannot
use the compatible energy formalism described in this paper on top of a scheme
which also accounts for the effects of the variable $h$.  Indeed, one nice
aspect of the compatible formalism is that it can be used on top of any scheme
for computing the pair-wise accelerations $a_{ij}$ -- it does not really matter
how you arrived at the acceleration, only that you then self-consistently
account for the discrete work done by those accelerations.  Additionally, one
can compatibly introduce the work done by any physics that creates an
acceleration on the SPH points in a similar manner.  In this way the compatible
discretization actually improves on the simplicity of adding new physics to an
SPH scheme.

There are however computational expense tradeoffs to be made.  We find that by
spending memory to hold the pair-wise accelerations for each interacting pair of
nodes that the computational run-time for a typical problem is only increased by
a few percent in our implementation.  However, the required memory use may not
be practical on all computer architectures, in which case it will be necessary
to compute the pair-wise accelerations twice, most likely resulting in a larger
run-time penalty than we see here.

We believe that the improvements demonstrated here are relevant for typical SPH
applications, such as modeling galaxy formation.  For instance, the important
physical process in the Noh problem (\S \ref{NohPlanar.sec} \& \ref{NohCyl.sec})
is the conversion of the kinetic energy due to a convergent inflow to thermal
energy via a strong shock -- this mirrors one of the dominant processes in the
formation of galaxies and galaxy clusters, where the kinetic energy due to the
gravitational infall of the gas is converted almost entirely to thermal energy
through strong shocks.  The Taylor-Sedov blastwave (\S \ref{SedovPlanar.sec})
consists of the introduction of a strong thermal energy source into an ambient
gas, much like the way the process of star formation in SPH modeling of galactic
evolution introduces thermal energy released by the absorption of radiation from
hot young stars and supernovae back into the simulated interstellar medium.
There is a potential lesson for such models though in the fact that we still
found it beneficial (in terms of the accuracy of the final solution) to smooth
the initial energy deposition in the Taylor-Sedov problem.  This issue may
warrant further investigation.

JMO would like to acknowledge many valuable discussions with Doug Miller.  We
would also like to acknowledge some useful suggestions from the referees which
have contributed to the clarity of this manuscript. This work was performed
under the auspices of the U.S. Department of Energy by the University of
California, Lawrence Livermore National Laboratory under contract
No. W-7405-Eng-48.


\end{document}